\documentclass{article}
\usepackage{dsfont}
\usepackage{bm}
\usepackage{subfigure}
\usepackage{enumitem}
\usepackage{multirow,verbatim}

\usepackage{amssymb, amsmath, amsthm,graphicx}
\usepackage{natbib}

\usepackage{algorithm}
\usepackage{algpseudocode}

\newtheorem{thm}{Theorem}
\newtheorem{dfn}{Definition}

\newcommand{\mG}{\mathcal G}
\newcommand{\mV}{\mathcal V}
\newcommand{\mE}{\mathcal E}

\newcommand{\diag}[1]{\mathrm{diag}(#1)}
\newcommand{\sign}[1]{\mathrm{sign}(#1)}
\newenvironment{rcases}
{\left.\begin{aligned}}
	{\end{aligned}\right\rbrace}

\begin{document}

% Title of paper
\title{$\ell_1$-Penalized Censored Gaussian Graphical Model}

\author{Luigi Augugliaro$^\ast$\\[4pt]
	\textit{Department of Economics, Business and Statistics, University of Palermo, Palermo, Italy}\\[2pt]
	{luigi.augugliaro@unipa.it}\\[4pt]
	Antonino Abbruzzo\\[4pt]
	\textit{Department of Economics, Business and Statistics, University of Palermo,Palermo, Italy}\\[4pt]
	Veronica Vinciotti\\[4pt]
	\textit{Department of Mathematics, Brunel University London, Uxbridge UB8 3PH, United Kingdom}}	

% Running headers of paper:
%\markboth{L. Augugliaro and others}{$\ell_1$-Penalized Censored Gaussian Graphical Model}
% First field is the short list of authors
% Second field is the short title of the paper

\maketitle
% Add a footnote for the corresponding author if one has been
% identified in the author list
%\footnotetext{To whom correspondence should be addressed.}

\begin{abstract}
Graphical lasso is one of the most used estimators for inferring genetic networks. Despite its diffusion, there are several fields in applied research where the limits of detection of modern measurement technologies make the use of this estimator theoretically unfounded, even when the assumption of a multivariate Gaussian distribution is satisfied. Typical examples are data generated by polymerase chain reactions and flow cytometer. The combination of censoring and high-dimensionality make inference of the underlying genetic networks from these data very challenging. In this paper we propose an $\ell_1$-penalized Gaussian graphical model for censored data and derive two EM-like algorithms for inference. By an extensive simulation study, we evaluate the computational efficiency of the proposed algorithms and show that our proposal overcomes existing competitors when censored data are available. We apply the proposed method to gene expression data coming from microfluidic RT-qPCR technology in order to make inference on the regulatory mechanisms of blood development.%

\end{abstract}

\section{Introduction\label{sec1}}

An important aim in genomics is to understand interactions among genes, characterized by the regulation and synthesis of proteins under internal and external signals. These relationships can be represented by a genetic network, i.e., a graph where nodes represent genes and edges describe  the interactions among them. Genetic networks can be used to gain new insights into the activity of biological pathways and to deduce unknown functions of genes from their dependence onto other genes. %Furthermore, incorporating this information into further statistical analyses has successfully resulted in the identification of novel biomarkers and more accurate classification methods (\citealp{YousefEtAl_BMCBioInfo_09, ChenEtAl_BMC_SystBio_11}).

Gaussian graphical models (\citealp{Dempster_BioS_72}) have been widely used  for reconstructing a genetic network from expression data. The reason of such diffusion relies on the statistical properties of the multivariate
Gaussian distribution which allow the topological structure of a network to be related with the non-zero elements of the concentration matrix, i.e., the inverse of the covariance matrix. Thus, the problem of network inference can be recast as the problem of estimating a concentration matrix. The graphical lasso (\citealp{YuanEtAl_BioK_07}) is a popular method for estimating a sparse concentration matrix, based on the idea of adding an $\ell_1$-penalty to the likelihood function of the multivariate Gaussian distribution. Nowadays, this estimator is widely used in applied research (e.g. \citealp{MenendezEtAl_Plos1_10, VinciottiEtAl_SABMB_16}) and widely studied in the computational (e.g., \citealp{FriedmanEtAl_biostat_08, WittenEtAl_JCGS_11, MazumderETAL_EJS_12}) as well as theoretical (e.g., \citealp{BickelEtAl_AOS_08a, GouEtAl_BioK_11}) literature. The interested reader is referred to \citet{AugugliaroEtAl_Book_2016} for an extensive  review.
%
%To address this inferential problem, in \citet{YuanEtAl_BioK_07} is proposed a new estimator, called graphical lasso (cglasso), defined by adding the $\ell_1$-penalty function to the density function. Nowadays, the cglasso estimator is widely used in applied research (e.g. \citealp{MenendezEtAl_Plos1_10, VinciottiEtAl_SABMB_16}) and it is one of the most active area of computational (e.g., \citealp{FriedmanEtAl_biostat_08, WittenEtAl_JCGS_11, MazumderETAL_EJS_12}) and theoretical (e.g., \citealp{BickelEtAl_AOS_08a, GouEtAl_BioK_11}) research. The interest reader is referred to \citet{AugugliaroEtAl_Book_2016} for an extensive  review.

Despite the widespread literature on the graphical lasso estimator, there is a great number of fields in applied research where modern measurement technologies make the use of this graphical model theoretically unfounded, even when the assumption of a multivariate Gaussian distribution is satisfied. A first example of this is Reverse Transcription quantitative Polymerase Chain Reaction (RT-qPCR), a popular technology for gene expression profiling (\citealp{DerveauxEtAl_Methods_10}). This technique relies on fluorescence-based detection of amplicon DNA and allows the kinetics of PCR amplification to be monitored in real time, making it possible to quantify nucleic acids with extraordinary ease and precision. The analysis of the raw RT-qPCR profiles is based on the cycle-threshold, defined as the fractional cycle number in the log-linear region of PCR amplification in which the reaction reaches fixed amounts of amplicon DNA (\citealp{GuesciniEtAl_BMSBioInfo_08}). If a target is not expressed or the amplification step fails, the threshold is not reached after the maximum number of cycles (limit of detection) and the corresponding cycle-threshold is undetermined. For this reason, the resulting data is naturally right-censored data (e.g., \citealp{PipelersEtAl_Plos1_17, McCallEtAl_BioInfo_14}). Another example is given by the flow cytometer, which is an essential tool in the diagnosis of diseases such as acute leukemias and malignant lymphomas (\citealp{BrownEtAl_ClinicChem_00}). A flow cytometer measures a limited range of signal strength and records each marker value within a fixed range, such as between 0 and 1023. If a measurement falls outside this range, then the value is replaced by the nearest legitimate value; that is, a value smaller than 0 is censored to 0 and a value larger than 1023 is censored to 1023. A direct application of the graphical lasso for network inference from data such as these is theoretically unfunded since it does not consider the effects of the censoring mechanism on the estimator of the concentration matrix.

In this paper we propose an extension of the graphical lasso estimator that  takes into account the censoring mechanism of the data  explicitly. Our work can be related to \cite{StadlerEtAl_StatComp_12}, who propose an $\ell_1$-penalized estimator of the inverse covariance matrix of a multivariate Gaussian model based on the assumption that the data are missing at random. As we shall see in the following of this paper, failure to take into account the censoring mechanism causes a poor behaviour of this approach when compared to our proposal. Our proposal can also be related to the work of \cite{PerkinsEtAl_AcadRadiol_13}, \cite{HoffmannEtAl_JABES_14} and \cite{PesonenEtAl_CSDA_15}, who provide a maximum likelihood estimator of the covariance matrix under left-censoring. However, these works do not address the estimation of the precision matrix under a sparsity assumption and for this reason they are applicable only when the sample size is larger than the number of nodes in the network.
%
%
%
%Another proposals that can be related to our method are in \cite{PerkinsEtAl_AcadRadiol_13}, \cite{HoffmannEtAl_JABES_14} and \cite{PesonenEtAl_CSDA_15}, where are studied the theoretical and computational aspects of the maximum likelihood estimator of the covarianca matrix when  left-censored data are available. In these works is not address the problem of how estimate the precision matrix under sparsity assumption, for this reason they are applicable only when the sample size is large enough.

%is in \cite{PerkinsEtAl_AcadRadiol_13} where is proposed a maximum likelihood estimator for the mean vector and the covariance matrix based on the multivariate normality assumption in the presence of left-censored data. \cite{HoffmannEtAl_JABES_14} proposed a pseudo-likelihood approach in order to overcome the complexity of estimating the parameters of multivariate Gaussian distribution while \cite{PesonenEtAl_CSDA_15} proposed a pairwise maximum likelihood estimation of the covariance matrix with an adjustment to achieve positive semi-definiteness of the resulting estimate. We underline that in all this works is not address the problem of how to estimate the precision matrix under sparsity assumption, for this reason they are applicable only when the sample size is large enough.

The remaining part of this paper is structured as follows. In Section~\ref{sec:cGGM}, we extend the notion of Gaussian graphical model to censored data and in Section~\ref{sec:cglasso} we propose the extension of the graphical lasso estimator to a censored graphical lasso estimator and two Expectation-Maximization (EM) algorithms  for inference of parameters in the censored Gaussian graphical model.  Section~\ref{sec:sim_study} is devoted to the evaluation of the behaviour of the proposed algorithms and the comparison of the proposed estimator with existing competitors. Finally, in Section~\ref{sec:realdata} we study a real dataset and in Section~\ref{sec:conclusions} we draw some conclusions.

\section{The censored Gaussian graphical model\label{sec:cGGM}}
%Let us briefly review the basic notions of graph theory and its relationship with the graphical models, in order to make this paper self-container.
In this section, we describe the technical details of the proposed method.
Let $\bm X = (X_1,\ldots,X_p)^\top$ be a $p$-dimensional random vector. Graphical models allow to represent the set of conditional independencies among these random variables by a graph $\mG = \{\mV,\mE\}$, where $\mV$ is the set of nodes associated to $\bm X$ and $\mE\subseteq\mV\times\mV$ is the set of ordered pairs, called edges, representing the conditional dependencies among the $p$ random variables (\citet{Lauritzen_book_96}, \citet{pearl1997graphical}, \citet{whittaker2009graphical}).

The Gaussian graphical model is a member of this class of models based on the assumption that  $\bm X$ follows a multivariate Gaussian distribution with expected value $\bm\mu = (\mu_1,\ldots,\mu_p)^\top$ and covariance matrix $\Sigma = (\sigma_{hk})$. Denoting with $\Theta = (\theta_{hk})$ the concentration matrix, i.e., the inverse of the covariance matrix, the density function of $\bm X$ can be written as follows
\begin{equation}\label{eqn:gauss_density}
\phi(\bm x;\bm\mu, \Theta) = (2\pi)^{-p/2}|\Theta|^{1/2}\exp\{-1/2 (\bm x - \bm\mu)^\top\Theta(\bm x -  \bm\mu) \}.
\end{equation}
As shown in \cite{Lauritzen_book_96}, the off-diagonal elements of the concentration matrix are the parametric tools relating the pairwise Markov property to the factorization of the density (\ref{eqn:gauss_density}). Formally, two random variables, say $X_h$ and $X_k$, are conditionally independent given all the remaining variables if and only if $\theta_{hk}$ is equal to zero. This result provides a simple way to relate the topological structure of the graph $\mG$ to the pairwise Markov property, i.e., the undirected edge $(h,k)$ is an element of the edge set $\mE$ if and only if $\theta_{hk}\ne 0$, consequently the graph specifying the factorization of the density (\ref{eqn:gauss_density}) is also called the concentration graph. Taking this into consideration, a Gaussian graphical model can be formally defined as the triple $\{\bm X, \phi(\bm x;\bm\mu, \Theta), \mG\}$ where the density $\phi(\bm x;\bm\mu, \Theta)$ factorizes according to the concentration graph $\mG$.%, meaning that $\theta_{hk} = 0$ if and only if $(h,k)\notin\mE$.

\begin{figure}[!p]
\centering
\includegraphics[scale=0.43]{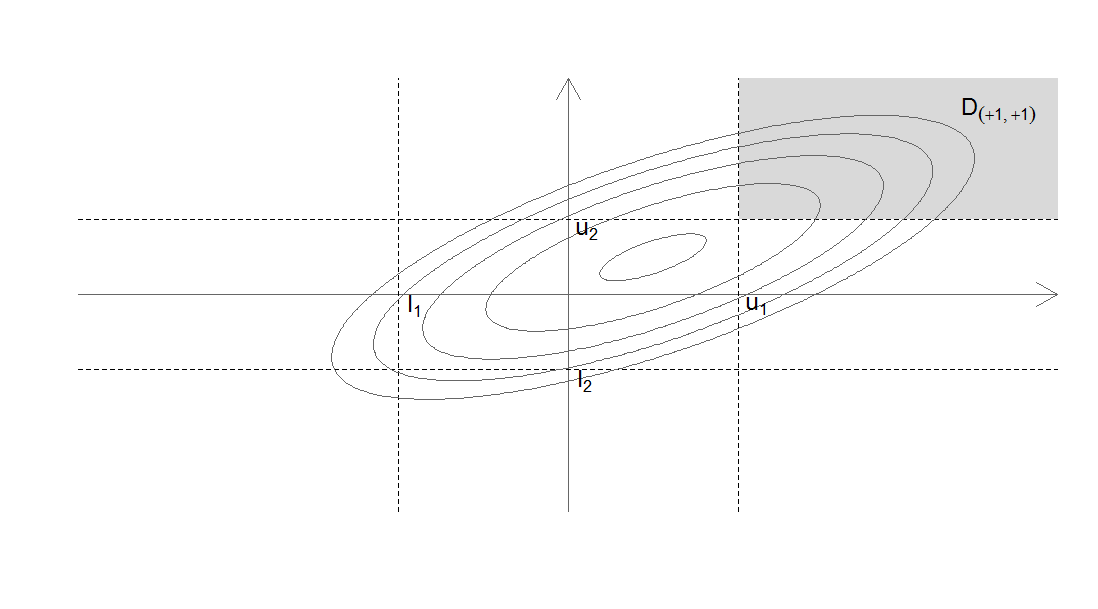}
\caption{Censoring mechanism for two random variables with upper and lower censoring vectors $\bm l = (l_1, l_2)^\top$ and $\bm u = (u_1, u_2)^\top$, respectively.}
\label{Figlikcens}
\end{figure}

Let $\bm X$ be a (partially) latent random vector with density function (\ref{eqn:gauss_density}). In order to include the censoring mechanism inside our framework, let us denote by $\bm l = (l_1,\ldots,l_p)^\top$ and $\bm u = (u_1,\ldots,u_p)^\top$, with $l_h < u_h$ for $h=1,\ldots,p$, the vectors of  known left and right censoring values. Thus, $X_h$ is observed only if it is inside the interval $[l_h,u_h]$ otherwise it is censored from below if $X_h<l_h$ or censored from above if $X_h > u_h$. Under this setting, a rigorous definition of the joint distribution of the observed data can be obtained using the approach for missing data with nonignorable mechanism (\citealp{LittleEtAl_Book_02}). This requires the specification of the distribution of a $p$-dimensional random vector, denoted by $R(\bm X;\bm l, \bm u)$, used to encode the censoring patterns. Formally, the $h$th element of $R(\bm X;\bm l, \bm u)$ is defined as $R(X_h; l_h, u_h) = I(X_h>u_h) - I(X_h < l_h)$, where $I(\cdot)$ denotes the indicator function. By construction $R(\bm X;\bm l, \bm u)$ is a discrete random vector with support in the set $\{-1,0,1\}^p$ and probability  function
\[
\mbox{Prob}\{R(\bm X;\bm l, \bm u) = \bm r\} = \int_{D_{\bm r}} \phi(\bm x;\bm\mu, \Theta)\text{d} \bm x,
\]
where  $D_{\bm r} = \{\bm x\in \mathds{R}^p : R(\bm x;\bm l,\bm u) = \bm r\}$.
Let us give a small example about the censoring mechanism in which we consider only two random variables, i.e. $\bm X = (X_1, X_2)^\top$, with censoring vectors $\bm l = (l_1, l_2)^\top$ and $\bm u = (u_1, u_2)^\top$. As shown in Figure~\ref{Figlikcens}, the censoring vectors entail a partition of the space $\mathds R^2$ in nine subsets, whereby each one of them is associated with a vector $\bm r\in\{-1,0,+1\}^2$ codifying a specific censoring pattern. In our example, the gray region is associated to the case where both random variables are right censored, i.e., $\bm r = (+1, +1)^\top$, and the probability to observe this pattern is obtained by integrating the bivariate normal density function over the region $D_{(+1, +1)} = \{\bm x\in \mathds R^2 : R(\bm x, \bm l, \bm r) = (+1, +1)^\top\}$, i.e., $\mbox{Prob}\{R(\bm X;\bm l, \bm u) = \bm r\} = \int_{D_{\bm r}} \phi(\bm x;\bm\mu, \Theta)\text{d} \bm x =  \int_{u_1}^{+\infty}\int_{u_2}^{+\infty} \phi(\bm x;\bm\mu, \Theta)\text{d}x_1\text{d}x_2$.

Given a censoring pattern, we can simplify our notation by partitioning the set $\mathcal I = \{1,\ldots,p\}$ into the sets $o=\{h\in\mathcal I:r_h = 0\}, c^-=\{h\in\mathcal I:r_h = -1\}$ and $c^+=\{h\in\mathcal I:r_h = +1\}$ and, in the following of this paper, we shall use the convention that a vector indexed by a set of indices denotes the corresponding subvector. For example, the subvector of observed elements in $\bm x$ is denoted by $\bm x_o = (x_h)_{h\in o}$ and, consequently, the observed data is the vector $(\bm x_o^\top, \bm r^\top)^\top$. As explained in \citet{LittleEtAl_Book_02}, the joint probability distribution of the observed data, denoted by $\varphi(\bm x_o,\bm r;\bm \mu,\Theta)$, is obtained by integrating $\bm X_{c^+}$ and $\bm X_{c^-}$ out of the joint distribution of $\bm X$ and $R(\bm X;\bm l, \bm u)$, which can be written as the product of the density function (\ref{eqn:gauss_density}) and the conditional distribution of $R(\bm X;\bm l, \bm u)$ given $\bm X = \bm x$. Formally
\begin{equation}\label{eqn:obs_density1}
\varphi(\bm x_o,\bm r;\bm \mu,\Theta) = \int\phi(\bm x_o, \bm x_{c^-},\bm x_{c^+};\bm\mu, \Theta) \mbox{Prob}\{R(\bm X;\bm l, \bm u) = \bm r \mid \bm X = \bm x\} \text{d}\bm x_{c^-}\text{d}\bm x_{c^+}.
\end{equation}
The density function (\ref{eqn:obs_density1}) can be further simplified by observing that $\mbox{Prob}\{R(\bm X;\bm l, \bm u) = \bm r \mid \bm X = \bm x\}$ is equal to one if the censoring pattern encoded in $\bm r$ is equal to the pattern observed in $\bm x$,  otherwise it is equal to zero, i.e.,
\[
\mbox{Prob}\{R(\bm X;\bm l, \bm u) = \bm r \mid \bm X = \bm x\} =  I(\bm x_{c^-} < \bm l_{c^-}) I(\bm l_o \le \bm x_o \le \bm u_o) I(\bm u_{c^+} < \bm x_{c^+}),
\]
where the inequalities in the previous expressions are intended elementwise. From this, $\varphi(\bm x_o,\bm r;\bm \mu,\Theta)$ can be rewritten as
\begin{eqnarray}\label{eqn:obs_density}
\varphi(\bm x_o, \bm r;\bm\mu, \Theta) &=& \int_{\bm u_{c^+}}^{+\infty}\int_{-\infty}^{\bm l_{c^-}}\phi(\bm x_o, \bm x_{c^-},x_{c^+};\bm\mu, \Theta) \text{d}\bm x_{c^-}\text{d}\bm x_{c^+} I(\bm l_o \le \bm x_o \le \bm u_o) \nonumber \\
&=& \int_{D_c} \phi(\bm x_o, \bm x_{c};\bm\mu, \Theta) \text{d}\bm x_{c} I(\bm l_o \le \bm x_o \le \bm u_o),
\end{eqnarray}
where $D_c = (-\infty, \bm l_{c^-})\times(\bm u_{c^+},+\infty)$ and $c = c^-\cup c^+$. The density function (\ref{eqn:obs_density}) is used in \citet{LeeEtAl_CSDA_12} inside the framework of a mixture of multivariate Gaussian distribution with censored data. Using (\ref{eqn:obs_density}) the censored Gaussian graphical model can be formally defined as follows.
\begin{dfn}\label{dfn:icggm}
Let $\bm X$ be a $p$-dimensional Gaussian distribution whose density function $\phi(\bm x;\bm\mu,\Theta)$ factorizes according to an undirected graph $\mG = \{\mV,\mE\}$ and let $R(\bm X;\bm l, \bm u)$ be a $p$-dimensional random censoring-data indicator defined by the censoring vectors $\bm l$ and $\bm u$. The censored Gaussian graphical model (cGGM) is defined to be the set $\{\bm X, R(\bm X;\bm l, \bm u), \varphi(\bm x_o, \bm r;\bm\mu, \Theta), \mG\}$.
\end{dfn}

A closer look at the Definition~\ref{dfn:icggm} reveals that  the proposed notion of censored Gaussian graphical model is characterized by a high degree of generality since it covers also the special cases of the classical Gaussian graphical model, obtained when $l_h = -\infty$ and $u_h = +\infty$ for $h=1,\ldots,p$, and the cases in which there is only  left-censored data ($u_h=+\infty$ for any $h$) or right-censored data ($l_h=-\infty$ for any $h$). %In the following of this paper a cGGM with only right-censored data will be called right-censored Gaussian graphical model (right-cGGM); the same for the left-censored Gaussian graphical model (left-cGGM).

\section{$\ell_1$-penalized Estimator for Censored Gaussian Graphical Model\label{sec:cglasso}}

\subsection{The censored graphical lasso estimator\label{subsec:cglasso}}

Consider a sample of $n$ independent observations drawn from the censored Gaussian graphical model $\{\bm X, R(\bm X;\bm l, \bm u), \varphi(\bm x_o, \bm r;\bm\mu, \Theta), \mG\}$. For ease of exposition, we shall assume that $\bm l$ and $\bm u$ are fixed across the $n$ observations, but  the extension to the cases where the censoring vectors are specific to each observation is straightforward and does not require a specific treatment.

As in the previous section, the set of indices of the variables observed in the $i$th observation is denoted by $o_i = \{h\in\mathcal I:r_{ih} = 0\}$, while $c^-_i=\{h\in\mathcal I:r_{ih} = -1\}$ and $c^+_i=\{h\in\mathcal I:r_{ih} = +1\}$ denote the sets of indices associated to the left and right-censored data, respectively. Denoting by $\bm r_i$ the $i$th realization of the random vector $R(\bm X_i;\bm l, \bm u)$, the $i$th observed data is the vector $(\bm x_{io_i}^\top, \bm r_i^\top)^\top$. Using the density function (\ref{eqn:obs_density}), the observed log-likelihood function can be written as
\begin{equation}\label{eqn:obssampleloglik}
\ell(\bm\mu,\Theta) = \sum_{i=1}^{n} \log\int_{D_{c_i}}\phi(\bm x_{io_i}, \bm x_{ic_i};\bm\mu, \Theta) \text{d} \bm x_{ic_i} = \sum_{i=1}^n\log\varphi(\bm x_{io_i},\bm r_i;\bm\mu,\Theta),
\end{equation}
where $D_{c_i} = (-\infty, \bm l_{c^-_i})\times (\bm u_{c^+_i},+\infty)$ and $c_i = c^-_i\cup c^+_i$. Although inference about the parameters of this model can be carried out via the maximum likelihood method, the application of this inferential procedure to real datasets, such as the gene expression data described in Section \ref{sec:realdata}, is limited for three main reasons. Firstly, the number of measured variables is often larger than the sample size and this implies the non-existence of the maximum likelihood estimator even when the dataset is fully observed. Secondly, even when the sample size is large enough, the maximum likelihood estimator will exhibit a very high variance (\citealp{SchaferEtAl_SAGMB_05, Uhler_AOS_12}). Thirdly, empirical evidence suggests that gene networks or more general biochemical networks are not fully connected (\citealp{GardnerEtAl_Science_03}). In terms of Gaussian graphical models this evidence translates in the assumption that $\Theta$ has a sparse structure, i.e., only few $\theta_{hk}$ are different from zero, which is not obtained by a direct (or indirect) maximization of the observed log-likelihood function (\ref{eqn:obssampleloglik}).

%To deal with high-dimensional data, researchers have been focused on penalized models which add a penalty term $P_\rho(\Theta)$ to the log-likelihood.
%The penalty term $P_\rho(\Theta)$ can be, for example, the LASSO \citep{FriedmanEtAl_biostat_08}, the ADAPTIVE LASSO \citep{zou2006adaptive},
%SCAD \citep{fan2001variable}, the ELASTIC NET \citep{zou2005regularization} and the FUSED LASSO \citep{tibshirani2005sparsity}. These methods produce sparsity relying on the $\ell_1$-norm.

All that considered, in this paper we propose to estimate the parameters of the censored Gaussian graphical model by generalizing the approach  proposed in \citet{YuanEtAl_BioK_07}, i.e., by maximizing a new objective function defined by adding a lasso-type penalty function to the observed log-likelihood (\ref{eqn:obssampleloglik}). The resulting estimator, called censored graphical lasso (cglasso), is formally defined as
\begin{equation}\label{dfn:cglasso}
\{\bm{\hat\mu}^\rho,\widehat\Theta^\rho\} = \arg\max_{\bm\mu,\Theta\succ0}\frac{1}{n} \sum_{i=1}^{n}\log \varphi(\bm x_{io_i},\bm r_i;\bm\mu,\Theta) - \rho \sum_{h\ne k}|\theta_{hk}|.
\end{equation}
Like in the standard graphical lasso estimator, the non-negative tuning parameter $\rho$ is used to control the amount of sparsity in the estimated concentration matrix $\widehat\Theta^\rho = (\hat\theta_{hk}^\rho)$ and, consequently, in the corresponding estimated concentration graph $\widehat\mG^\rho = \{\mV, \widehat\mE^\rho\}$, where $\widehat\mE^\rho = \{(h,k):\hat\theta^\rho_{hk} \ne0\}$. When $\rho$ is large enough, some  $\hat\theta_{hk}^\rho$ are shrunken to zero resulting in the removal of the corresponding link in $\widehat\mG^\rho$; on the other hand, when $\rho$ is equal to zero and the sample size is large enough the estimator $\widehat\Theta^\rho$ coincides with the maximum likelihood estimator of the concentration matrix, which implies a fully connected estimated concentration graph.

\subsection{Fitting the censored graphical lasso model\label{sec:algorirhm_cglasso}}

%Before to deal with the technical details of the algorithm proposed to compute the cglasso estimator, we start this section with the necessary notation.
Using known results about the multivariate Gaussian distribution, the conditional distribution of $\bm X_{c_i}$ given $\bm X_{o_i} = \bm x_{o_i}$ is also a multivariate Gaussian distribution with concentration matrix $\Theta_{c_ic_i} = (\theta_{hk})_{h,k\in c_i}$ and conditional expected value equal to $E_{c_i\mid o_i}(\bm X_{c_i}) = \bm\mu_{c_i\mid o_i} = \bm\mu_{c_i} - \Theta^{-1}_{c_ic_i}\Theta_{c_io_i}(\bm x_{io_i} - \bm\mu_{o_i})$, where $\Theta_{c_io_i} = (\theta_{hk})_{h\in c_i,k\in o_i}$. As we shall show in the next theorem, the conditions characterizing the cglasso estimator are based on the the first and second moment of the Gaussian distribution $\phi(\bm x_{c_i};\bm\mu_{c_i\mid o_i}, \Theta_{c_ic_i})$ truncated over the region $D_{c_i}$. To this end we let
\[
\begin{array}{rl}
x_{i,h}(\bm\mu,\Theta) & = \left\{%
\begin{array}{ll}
x_{ih} & \mbox{ if } r_{ih} = 0\\
E_{c_i\mid o_i}(X_{ih}\mid \bm X_{ic_i}\in D_{c_i}) & \mbox{ otherwise},
\end{array}
\right.%
\\
\\
x_{i,hk}(\bm\mu,\Theta) & = \left\{%
\begin{array}{ll}
x_{ih} x_{ik}& \mbox{ if } r_{ih} = 0 \mbox { and } r_{ik} = 0\\
x_{ih}E_{c_i\mid o_i}(X_{ik}\mid \bm X_{ic_i}\in D_{c_i}) & \mbox{ if } r_{ih} = 0 \mbox { and } r_{ik} \ne 0\\
E_{c_i\mid o_i}(X_{ih}\mid \bm X_{ic_i}\in D_{c_i}) x_{ik}& \mbox{ if } r_{ih} \ne 0 \mbox { and } r_{ik} = 0\\
E_{c_i\mid o_i}(X_{ih}X_{ik}\mid \bm X_{ic_i}\in D_{c_i}) & \mbox{ if } r_{ih} \ne 0 \mbox { and } r_{ik} \ne 0,\\
\end{array}
\right.
\end{array}
\]
where $E_{c_i\mid o_i}(\cdot\mid \bm X_{ic_i}\in D_{c_i})$ denotes the expected value computed using the conditional distribution of $\bm X_{ic_i}$ given $\bm x_{io_i}$ truncated over $D_{c_i}$. Finally we let $\bar x_h(\bm{\mu},\Theta) = \sum_{i=1}^n x_{i,h}(\bm\mu,\Theta) / n$, $\bm{\bar x}(\bm{\mu},\Theta) = \{\bar x_1(\bm{\mu},\Theta),\ldots,\bar x_p(\bm{\mu},\Theta)\}^\top$, $s_{hk}(\bm\mu,\Theta) = \sum_{i=1}^n x_{i,hk}(\bm\mu,\Theta) / n - \bar x_h(\bm\mu,\Theta)\bar x_k(\bm\mu,\Theta)$ and $S(\bm\mu,\Theta) = \{s_{hk}(\bm\mu,\Theta)\}$. Using this notation, the next theorem gives the Karush-Kuhn-Tuker conditions for the proposed estimator.

\begin{thm}\label{thm:kkt_cglasso}
Necessary and sufficient conditions for $\{\bm{\hat\mu}^\rho,\widehat\Theta^\rho\}$ to be the solution of the maximization problem
\[
\max_{\bm\mu,\Theta\succ0}\frac{1}{n} \sum_{i=1}^{n}\log \varphi(\bm x_{io_i},\bm r_i;\bm\mu,\Theta) - \rho \sum_{h\ne k}|\theta_{hk}|,
\]
are
\begin{equation}
\begin{rcases}
\bar x_h(\bm{\hat\mu}^\rho,\widehat\Theta^\rho) - \hat\mu_h^\rho &=  0  \\
\hat\sigma^\rho_{hk}(\bm{\hat\mu}^\rho,\widehat\Theta^\rho) - s_{hk}(\bm{\hat\mu}^\rho,\widehat\Theta^\rho) - \rho \hat v_{hk} &= 0
\end{rcases}
\label{eqn:kkt_1}
\end{equation}
where $\hat v_{hk}$ denotes the subgradient of the absolute value function at $\hat\theta^\rho_{hk}$, i.e., $\hat v_{hk} = \sign{\hat\theta^\rho_{hk}}$ if $\hat\theta^\rho_{hk}\ne0$ and $|\hat v_{hk}|\le 1$ if $\hat\theta^\rho_{hk}=0$.
\end{thm}
A proof of this theorem is reported in the supplementary material available at \textit{Biostatistics} online. The stationary conditions (\ref{eqn:kkt_1}) show that, while in the standard graphical lasso the parameter $\bm\mu$ can be estimated by the empirical average regardless of $\rho$ and the inference about the concentration matrix can be carried out using the profile log-likelihood function, inside our framework the two inferential problems cannot be separated, since the tuning parameter also affects the estimator of the expected value. Furthermore, the conditions suggest that, for a given value of the tuning parameter, the cglasso estimator can be computed using the EM algorithm \citep{DempsterEtAl_JRSSB_77}. This algorithm is based on the idea of repeating two steps until a convergence criterion is met. The first step, called E-Step, requires the calculation of the conditional expected value of the complete log-likelihood function using the current estimates. The resulting function, called $Q$-function, is maximized in the second step, i.e., the M-Step. As explained in \cite{McLachlanEtAl_Book_08}, the EM algorithm can be significantly simplified when the complete probability density function is a member of the regular exponential family. In this case the E-Step simply requires the computation of the conditional expected values of the sufficient statistics. In our case, if we denote by $\{\bm{\hat\mu}^\rho_{ini},\widehat\Theta^\rho_{ini}\}$ an initial estimate of the parameters, the E-Step reduces to the computation of $x_{i,h}(\bm{\hat\mu}^\rho_{ini},\widehat\Theta^\rho_{ini})$ and $x_{i,hk}(\bm{\hat\mu}^\rho_{ini},\widehat\Theta^\rho_{ini})$, for $i=1,\ldots, n$. From this, conditions (\ref{eqn:kkt_1}) can be written as
\begin{equation}
\begin{rcases}
\bar x_h(\bm{\hat\mu}^\rho_{ini},\widehat\Theta^\rho_{ini}) - \hat\mu_h^\rho &=  0 \\
\hat\sigma^\rho_{hk}(\bm{\hat\mu},\widehat\Theta) - s_{hk}(\bm{\hat\mu}^\rho_{ini},\widehat\Theta^\rho_{ini}) - \rho\hat v_{hk} &= 0
\end{rcases}
\label{eqn:kkt_2}
\end{equation}
which are the stationary conditions of a standard graphical lasso problem (\citealp{WittenEtAl_JCGS_11, MazumderETAL_EJS_12}) with $S(\bm{\hat\mu}^\rho_{ini},\widehat\Theta^\rho_{ini})$ used as the current estimate of the empirical covariance matrix. The conditions (\ref{eqn:kkt_2}) imply that in the M-Step the parameter $\bm\mu$ is estimated by $\bm{\bar x}(\bm{\hat\mu}^\rho_{ini},\widehat\Theta^\rho_{ini})$ while $\Theta$ is estimated by maximizing the following objective function
\begin{equation}\label{eqn:Qfun}
Q(\Theta\mid\widehat\Theta^\rho_{ini}) = \log\det\Theta - \mathrm{tr}\{\Theta S(\bm{\hat\mu}^\rho_{ini},\widehat\Theta^\rho_{ini})\} - \rho\sum_{h,k}|\theta_{hk}|.
\end{equation}

Table~\ref{algo:cglasso1} reports the pseudo-code of the proposed EM algorithm for the derivation of the cglasso estimator. Although the maximization of the $Q$-function (\ref{eqn:Qfun}) can be efficiently solved using, for example, the algorithm proposed by \citet{FriedmanEtAl_biostat_08}, by \citet{RothmanEtAl_EJS08} or by \cite{WittenEtAl_JCGS_11}, Table~\ref{algo:cglasso1} reveals that  the main computational cost of the proposed EM algorithm comes from the evaluation of the moments of the multivariate truncated Gaussian distribution, which are needed to compute $\bm{\bar x}(\bm{\hat\mu}^\rho_{ini},\widehat\Theta^\rho_{ini})$ and  $S(\bm{\hat\mu}^\rho_{ini},\widehat\Theta^\rho_{ini})$. These moments, which can be computed using the methods proposed by \citet{Tallis_JRSSB_61}, \citet{Lee_EL_83}, \citet{ LeppardEtAl_JRSSC_89} or by \citet{Arimsendi_JMA_13}, require complex numerical algorithms for the calculation of the integral of the multivariate normal density function (see \citet{GenzEtAl_JCGS_02} for a review), and they are computationally infeasible also for moderate size problems. For this reason, we also propose an approximated EM algorithm. In particular, using the mean field theory and following the idea proposed by \citet{GuoEtAl_JCGS_15}, we approximate the quantities $E_{c_i\mid o_i}(X_{ih}X_{ik}\mid \bm X_{ic_i}\in D_{c_i})$, for any $h\ne k$, by:
\begin{equation}\label{eqn:approxEXhXk}
E_{c_i\mid o_i}(X_{ih}X_{ik}\mid \bm X_{ic_i}\in D_{c_i}) \approx E_{c_i\mid o_i}(X_{ih}\mid \bm X_{ic_i}\in D_{c_i})E_{c_i\mid o_i}(X_{ik}\mid \bm X_{ic_i}\in D_{c_i}).
\end{equation}
As shown by \citet{GuoEtAl_JCGS_15}, the approach works well when the real concentration matrix is sparse or the tuning parameter is sufficiently large.
The main advantage coming from the approximation (\ref{eqn:approxEXhXk}) is that, to evaluate the quantities $\bm{\bar x}(\bm{\hat\mu}^\rho_{ini},\widehat\Theta^\rho_{ini})$ and  $S(\bm{\hat\mu}^\rho_{ini},\widehat\Theta^\rho_{ini})$, we now only need to compute $E_{c_i\mid o_i}(X_{ih}\mid \bm X_{ic_i}\in D_{c_i})$ and $E_{c_i\mid o_i}(X^2_{ih}\mid \bm X_{ic_i}\in D_{c_i})$, for $h=1,\ldots,p$, which can be computed by using exact formulas (\citealp{JohnsonEtAl_Book_94}) requiring only the evaluation of the cumulative distribution function of the univariate Gaussian distribution. In the following of this paper, by $\bar S(\bm{\hat\mu}^\rho_{ini},\widehat\Theta^\rho_{ini})$ we denote the approximated current estimate of the empirical covariance matrix. By an extensive simulation study, in Section~\ref{sec:sim_algo}  we shall study the behaviour of the Algorithm~\ref{algo:cglasso1} based on the usage of the matrix $\bar S(\bm{\hat\mu}^\rho_{ini},\widehat\Theta^\rho_{ini})$ in step 3 and 5.

As with graphical lasso, our proposed method requires a sequence of $\rho$-values, which should be suitably defined so as to reduce the computational cost needed to compute the entire path of the estimated parameters. The next theorem gives the exact formula for the derivation of the largest $\rho$-value, denoted by $\rho_{\max}$, and the corresponding cglasso estimator.
\begin{thm}\label{thm:maxrho}
For any index $h$ define the sets $o_h = \{i : r_{ih} = 0\}$, $c^{-}_h = \{i : r_{ih} = -1\}$ and $c^{+}_h = \{i : r_{ih} = +1\}$ and let
\begin{equation*}
\{\hat\mu_h,\hat\sigma^2_h\} = \arg\max_{\mu_h,\sigma^2_h}\sum_{i\in o_h} \log\phi(x_{ih};\mu_h,\sigma^2_h) + |c^{-}_h|\int_{-\infty}^{l_h} \phi(x;\mu_h,\sigma^2_h) \text{d} x + |c^{+}_h|\int_{u_h}^{+\infty} \phi(x;\mu_h,\sigma^2_h) \text{d} x.
\end{equation*}
Then the cglasso estimators corresponding to the largest value of the tuning parameter are equal to $\bm{\hat\mu}^{\rho_{\max}} = (\hat\mu_1,\ldots,\hat\mu_p)^\top$ and $\widehat\Theta^{\rho_{\max}} = \diag{\hat\sigma^{-2}_1,\ldots,\hat\sigma^{-2}_p}$ and $\rho_{\max} = \max_{h\ne k} |s_{hk}(\bm{\hat\mu}^{\rho_{\max}}, \widehat\Theta^{\rho_{\max}})|$.
\end{thm}

A proof of this theorem is reported in the supplementary material available at \textit{Biostatistics} online.  Using the results given in Theorem~\ref{thm:maxrho}, the entire path of the cglasso estimates can be computed as described in Algorithm~\ref{algo:cglasso2}, i.e., using the estimates obtained for a given $\rho$-value as warm starts for fitting the next cglasso model. This strategy is commonly used also in other efficient lasso algorithms and R packages (\citealp{FriedmanEtAl_JSS_10}, \citealp{glasso_manual}). In our model, this strategy turns out to be remarkably efficient since using a sufficiently fine sequence of $\rho$-values, the starting values defined in Step~5 of the Algorithm~\ref{algo:cglasso2} will be sufficiently close to the estimates computed in Step~6, thus removing the lack of convergence of the EM algorithms due to an inadequate choice of starting point (\citealp{McLachlanEtAl_Book_08}).

\subsection{Tuning parameter selection\label{sec:rhosel}}

The tuning parameter plays a central role in the proposed cglasso estimator since it is designed to control the complexity of the topological structure of the estimated concentration graph. In this paper we propose to select the optimal $\rho$-value of the cglasso estimator by using the Bayesian Information Criterion (BIC) which has successfully been applied in \citet{YuanEtAl_BioK_07} and references therein. For the proposed estimator, the BIC measure can be written as follows:
\[
\hbox{BIC}(\rho) = -\frac{2}{n}\sum_{i=1}^{n}\log \varphi(\bm x_{io_i},\bm r_i;\bm{\hat\mu}^\rho,\widehat\Theta^\rho) + \{2p + a(\rho)\}\frac{\log(n)}{n},
\]
where $a(\rho)$ denotes the number of nonzero off-diagonal estimates of $\widehat\Theta^\rho$. Although a grid search can be performed to select the $\rho$-value minimizing the $\hbox{BIC}(\rho)$ measure, the computational burden related to the evaluation of the log-likelihood function can make this strategy infeasible also for moderate size problems. For this reason, following the approach proposed in \citet{IbrahimEtAl_JASA_08}, we also propose to select the $\rho$-value by minimizing the following approximate BIC measure:
\[
\overline{\hbox{BIC}}(\rho) = -\log\det\widehat{\Theta}^\rho + \mathrm{tr}\{\Theta S(\bm{\hat\mu}^\rho,\widehat\Theta^\rho)\} + \{2p + a(\rho)\}\frac{\log(n)}{n},
\]
which is defined by substituting the exact log-likelihood function with the $Q$-function used in the M-Step of the proposed algorithm and is easily obtained as a byproduct of the EM algorithm. %The clear advantage to using $\overline{\hbox{BIC}}(\rho)$ instead of $\hbox{BIC}(\rho)$ is that the value of the $Q$-function at the solution point can be easily obtained as a byproduct of the EM algorithm.

\section{Simulation studies\label{sec:sim_study}}

\subsection{Evaluating the behaviour of the approximate EM algorithm\label{sec:sim_algo}}
In a first simulation, we evaluate the effects of the approximation (\ref{eqn:approxEXhXk}) on the accuracy of the estimators obtained with the EM algorithm. As in the real dataset studied in Section~\ref{sec:realdata}, we consider right-censored data and we set the right-censoring value $u_h$ equal to 40, for any $h = 1, \ldots, p$.  For this simulation, where we plan to use the full EM algorithm, we set the number of variables $p$ to 10 and the sample size $n$ to 100. To simulate a sample from a sparse right-censored model we use the following procedure. First, using the method implemented in the \texttt{R} package \texttt{huge} (\citealp{huge_manual}), we simulate a sparse concentration matrix with random structure, where we set the probability that a $\theta_{hk}$ is different from zero to 0.1. Then, the elements of the parameter $\bm\mu$ are selected to obtain a fixed probability of right censoring for any given random variable. More specifically, we randomly draw a subset $\mathcal D$ from $\mathcal I = \{1,\ldots,p\}$ and for each $h\in\mathcal D$ the value of the parameter $\mu_h$ is such that $\hbox{Prob}\{R(X_h;-\infty, 40) = +1\} = 0.25$ while for each $h\notin\mathcal D$ the parameter $\mu_h$ is such that the probability of right censoring is approximately equal to $10^{-11}$. In our study the cardinality of the set $\mathcal D$, denoted by $|\mathcal D|$, is used as a tool to analyze the effects of the number of censored variables on the behaviour of the Algorithm~\ref{algo:cglasso2} and its approximated version. Finally, we draw a sample from a multivariate Gaussian distribution with parameters given in the previous steps and we treat each value greater than 40 as a missing (censored) value. We simulated 100 datasets from this model and for each simulation we compute a path of cglasso estimates using the proposed EM algorithm and a path using the approximated EM algorithm. In the following of this section, the estimates belonging to the two paths are denoted by $\{\bm{\hat\mu}_e^\rho; \hat\Theta^\rho_e\}$ and $\{\bm{\hat\mu}_a^\rho; \hat\Theta^\rho_a\}$, respectively. For each path, the largest value of the tuning parameter was computed using the results given in Theorem~\ref{thm:maxrho} while the smallest value was set equal to $1\times10^{-3}$.

\begin{figure}[!p]
\centering
\subfigure[]
{\includegraphics[scale=0.4]{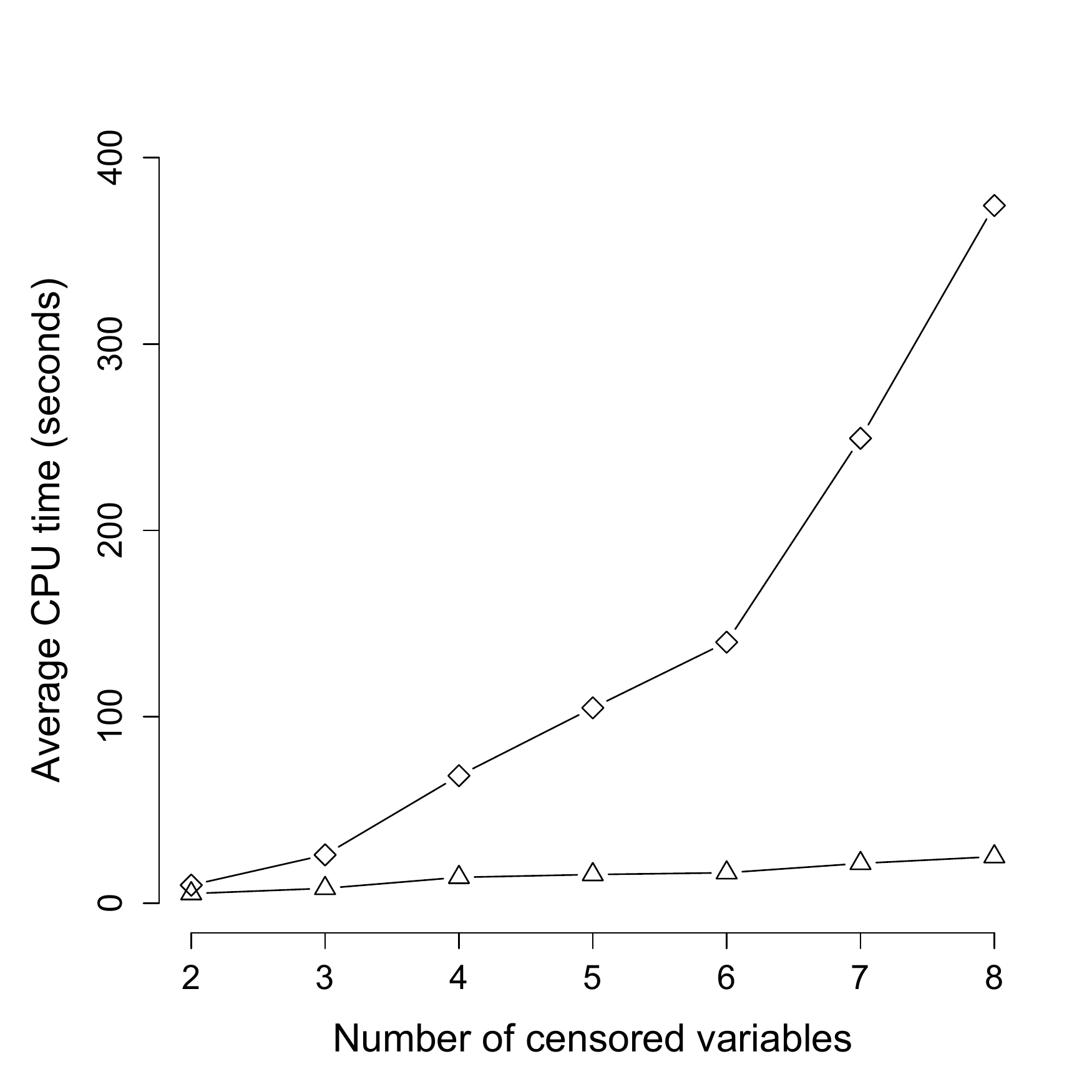}}
\subfigure[]
{\includegraphics[scale=0.4]{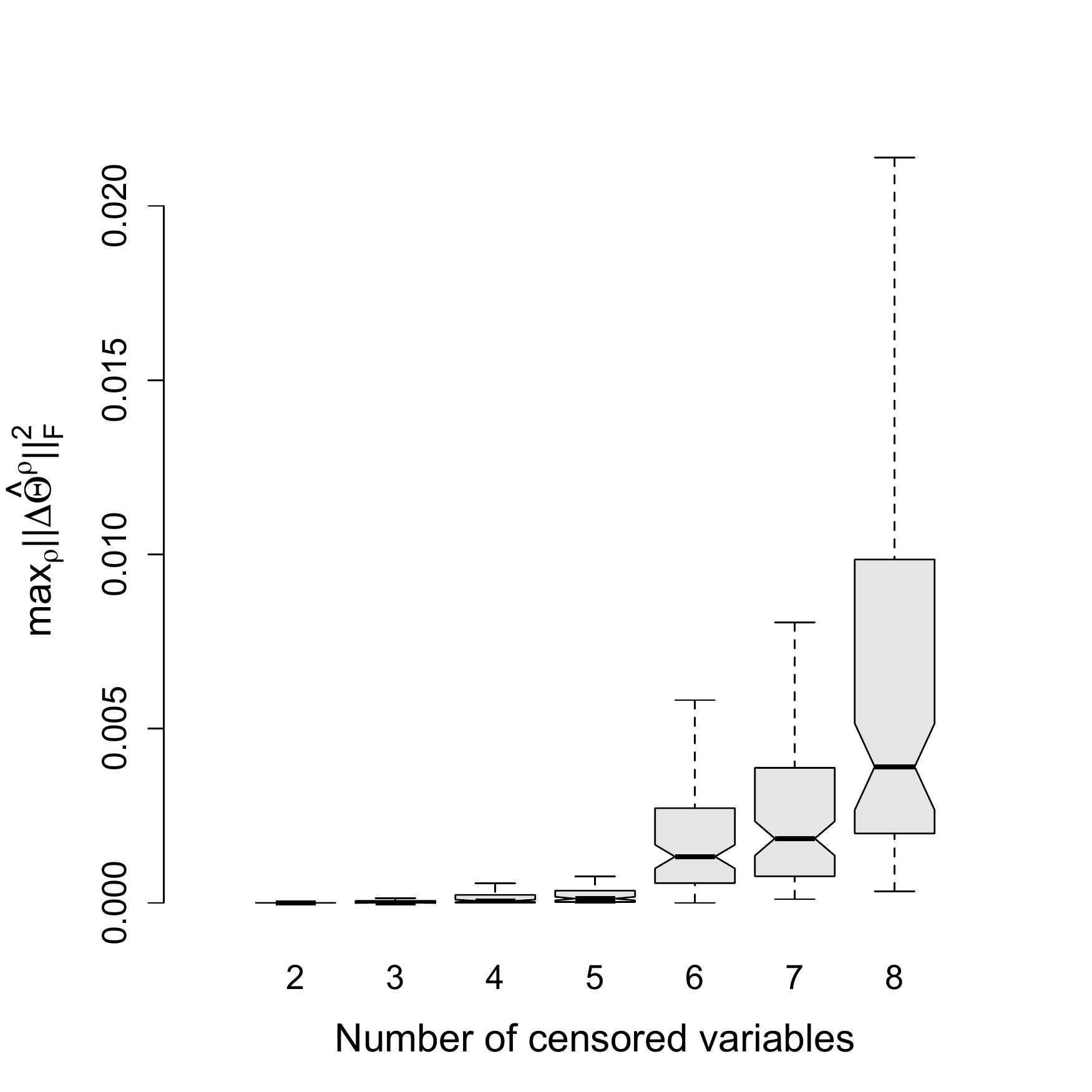}}
\caption{Panel~(a): average CPU time as a function of the cardinality of the set $\mathcal D$. Triangles and diamonds are refereed to results obtained by using the exact and approximated EM algorithm, respectively. Panel~(b):~box-plots of the largest Frobenius distance $\hat\Theta^\rho_e$ and $\hat\Theta^\rho_a$.}
\label{fig:CompEff}
\end{figure}

Panel (a) in Figure~\ref{fig:CompEff} shows the average CPU times for computing the path, as a function of the cardinality of the set $\mathcal D$. As expected, the computational time needed to compute a path is always an increasing function of the number of the censored variables but, when we use the matrix $S(\bm{\hat\mu}^\rho_{ini},\widehat\Theta^\rho_{ini})$ in the E-step, the figure reveals that the form of the relationship between the CPU time and $|\mathcal D|$ is almost exponential implying that the computation of the cglasso estimator is infeasible also for relatively small datasets. In contrast to this, when we use the matrix $\bar S(\bm{\hat\mu}^\rho_{ini},\widehat\Theta^\rho_{ini})$ to approximate the current estimate of the empirical covariance matrix, the figure shows an almost linear dependence of the CPU time, which implies a significant reduction of the computational complexity compared to the full EM algorithm. Although the results showed in Panel (a) strongly suggest the use of the approximated EM algorithm to compute the cglasso estimator, they do not provide information about the difference between the two estimates. For this reason, we also computed the largest Euclidean distance between $\bm{\hat\mu}_e^\rho$ and $\bm{\hat\mu}_a^\rho$, denoted as $\max_{\rho}\|\Delta\bm{\hat\mu}^\rho\|^2$, and the largest Frobenius distance between $\hat\Theta^\rho_e$ and $\hat\Theta^\rho_a$, denoted by $\max_{\rho}\|\Delta\hat\Theta^\rho\|^2_F$. Table~\ref{tbl:CompEff} reports the complete results, i.e., the average values and the standard deviations and  Panel~(b) shows the results for the $\max_{\rho}\|\Delta\hat\Theta^\rho\|^2_F$. All the results clearly show that the two estimators are sufficiently close to each other, and point to the use of the approximated EM algorithm for the derivation of the cglasso estimator.

\subsection{Comparison of methods on data simulated from a censored Gaussian graphical model}

\begin{figure}[!p]
\centering
\subfigure[]
{\includegraphics[scale=0.4]{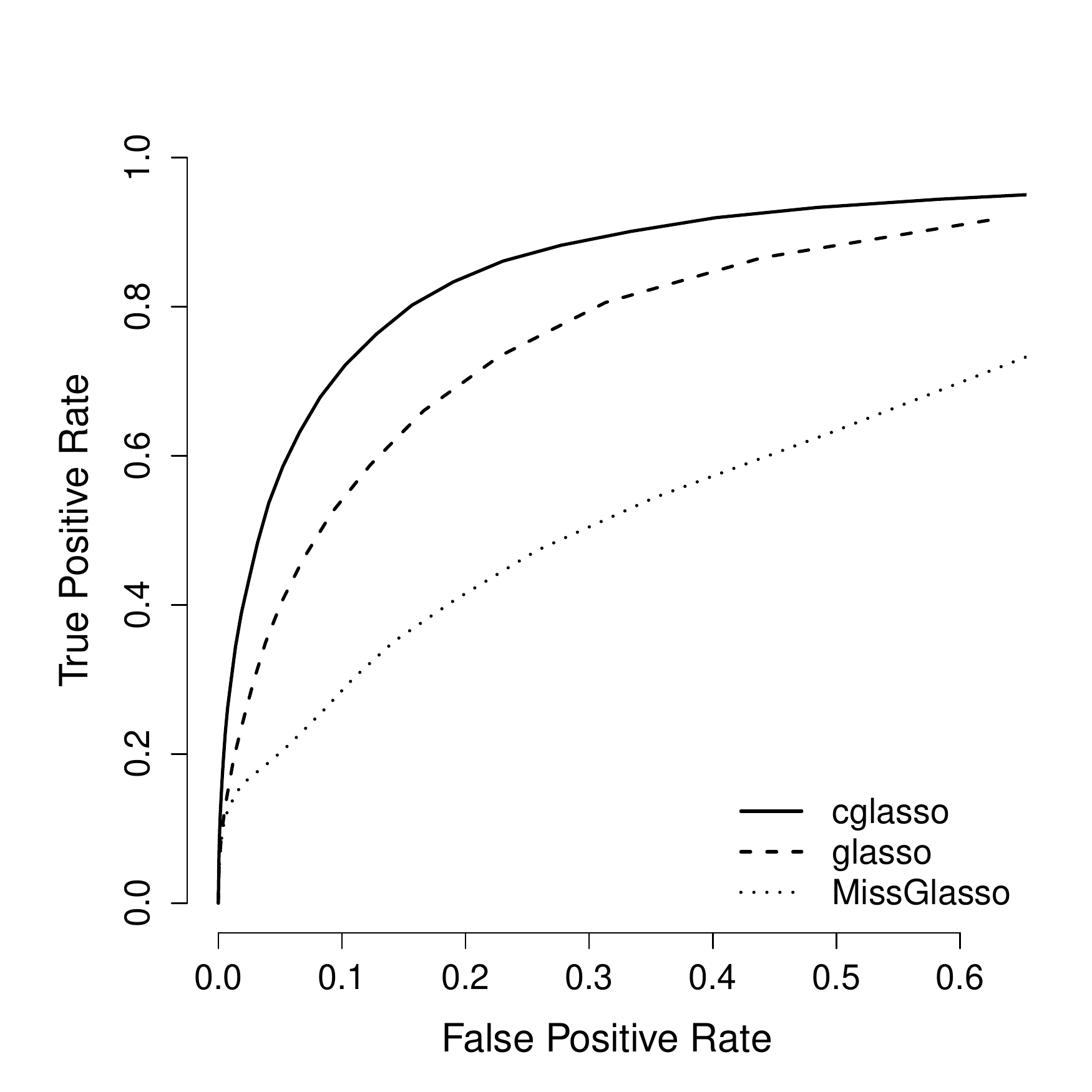}}
\subfigure[]
{\includegraphics[scale=0.4]{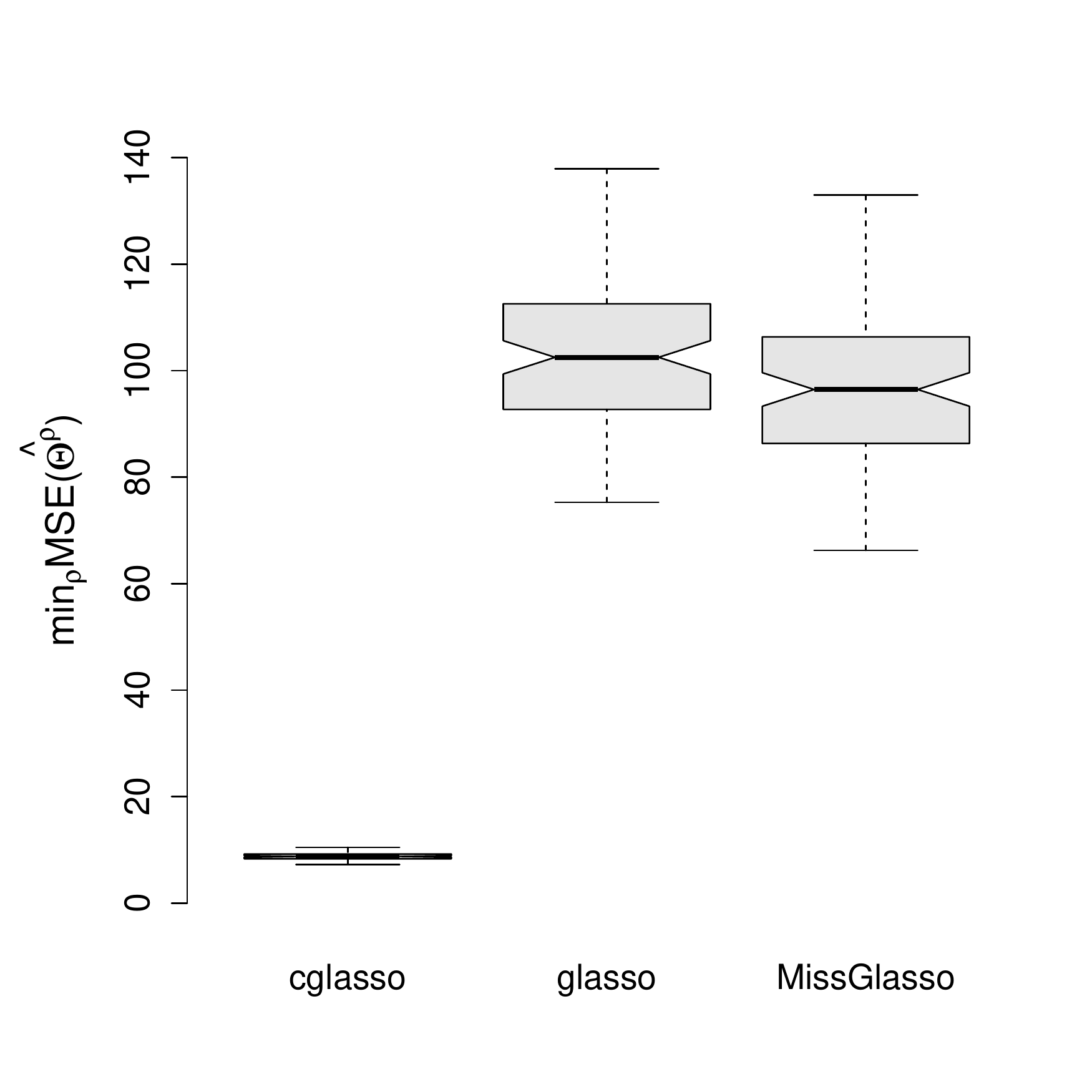}}
\caption{Results of the simulation study based on the Model 1 with $H = 25$. Panel (a) shows the ROC curves; Panel~(b) shows the box-plots of the beahviour of quantity $\min_{\rho}\hbox{MSE}(\widehat{\Theta}^\rho)$ for the considered estimators.}
\label{fig:simul_model1_a}
\end{figure}

In a second simulation study, we compare our proposed estimator with MissGlasso (\citet{StadlerEtAl_StatComp_12}), which performs $\ell_1-$penalised estimation under the assumption that the censored data are missing at random, and with the glasso estimator (\citealp{FriedmanEtAl_biostat_08}), where the empirical covariance matrix is calculated by imputing the missing values with the limit of detection. These estimators are evaluated in terms of both recovering the structure of the true concentration graph and the mean squared error.  We use a similar approach to the previous simulation for generating right censored data. In particular, we set the right censoring value to 40 for any variable and the sample size $n$ to 100. We generate a sparse concentration matrix with random structure and set the probability of observing a link between two nodes to $k/p$, where $p$ is the number of variables and $k$ is used to control the amount of sparsity in $\Theta$. Finally, we set the mean $\bm\mu$ in such a way that  $\mu_h = 40$ for the $H$ censored variables, i.e. $\hbox{Prob}\{R(X_h;-\infty, 40) = +1\} = 0.50$, while for the remaining variables $\mu_h$ is sampled from a uniform distribution on the interval $[10; 35]$. At this point, we simulate a sample from the latent $p$-variate Gaussian distribution and treat all values greater than 40 as missing. The quantities $k$, $p$ and $H$ are used to specify the different models used to analyze the behaviour of the considered estimators. In particular, we consider the following cases:
\begin{itemize}
\item\textbf{Model 1}: $k = 3$, $p = 50$ and $H \in\{25, 35\}$. The aim is to evaluate the effects of the number of censored variables on the behaviour of the proposed estimators in a setting in which $n>p$.
\item\textbf{Model 2}: $k \in\{1, 5\}$, $p = 50$ and $H  = 30$.  This setting is used to evaluate the effects of the sparsity of the matrix $\Theta$ on the considered estimators when $n > p$.
\item\textbf{Model 3}: $k = 3$, $p = 200$ and $H = 100$. This setting is used to evaluate the impact of the high dimensionality on the estimators ($p\gg n$).
\end{itemize}
For each model, we simulate 100 samples from the right censored model and in each simulation we compute the coefficients path using cglasso, MissGlasso and glasso.  Each path is computed using an equally spaced sequence of 30 $\rho$-values. Table~\ref{tbl:ResSimul} reports the summary statistics. The first meta-column shows the details of the models, the second and third columns  evaluate the statistical properties of the considered estimators in terms of minimum value of the mean squared error attained along a path of solutions and the Area Under the Curve (AUC) for network discovery, respectively. Note that the considered measures allow us to study the behaviour of the estimators along the entire path. %Instead, we could have focused on the behaviour of the estimators evaluated at a single point of the path selected by an adequate measure of goodness-of.fit, such as AIC or BIC. However, this approach might have a significant effect on the results.
The results on the AUC suggest that cglasso can be used as an efficient tool for recovering the structure of the true concentration matrix of a censored graphical model. The distribution of the minimum value of the mean squared errors, denoted by $\min_\rho\hbox{MSE}(\bm{\hat\mu}^\rho)$ and $\min_\rho\hbox{MSE}(\widehat\Theta^\rho)$, shows that, not only our estimator is able to recover the structure of the graph but also outperforms the competitors in terms of both estimation of $\bm\mu$ and $\Theta$. We did not report $\min_\rho\hbox{MSE}(\bm{\hat\mu}^\rho)$ for glasso since this method does not allow to estimate the parameter $\bm\mu$. Figure~\ref{fig:simul_model1_a} shows a graphical representation of the results for Model 1 with $H = 25$.

Panel~(a) in Figure~\ref{fig:simul_model1_a} shows the ROC curves. For any given value of the tuning parameter, cglasso gives a better estimate of the concentration matrix both in terms of true and false positive rate. The same behaviour was also observed in the other models used in our numerical study, and can be found in the Supplementary Material available at \textit{Biostatistics} online. Panel~(b) shows the distributions of the quantity $\min_\rho\hbox{MSE}(\widehat\Theta^\rho)$. These box-plots emphasize that, not only the cglasso has a mean squared error much smaller than glasso and MissGlasso, but also that it is much more stable than the competitors.

\subsection{Testing the robustness of the method on more realistic biological data}
In a third simulation study, we test the robustness of the method on more realistic biological data. In particular, we consider expression data from \cite{wille2004sparse} on the extensively studied Arabidopsis thaliana biological system. The study reports data from $n=118$ experiments on $p=39$ genes, whose regulatory network is of interest. Although the data are fully observed,  we test our method on an dataset where observations are made artificially censored, similarly to \cite{StadlerEtAl_StatComp_12}. In particular, we produce three datasets at different levels of right censoring, by recording as missing the 10\%, 20\% and 30\% of the highest values, respectively. Then, we compare our method, {\tt cglasso}, with {\tt MissGlasso} \citep{StadlerEtAl_StatComp_12} and with two additional methods which impute missing values first and then infer the network using graphical lasso on the imputed values. As in \cite{StadlerEtAl_StatComp_12}, we consider k-nearest neighbour imputation, which we denote with {\tt missknn}, and imputation using random forests, which we denote with {\tt MissForest}.
%First of all we scaled the data to produce a dataset with zero mean and unit variance. Based on the scaled data we created three datasets by deleting the 10\%, 20\% and 30\% of the highest values, respectively.

\begin{figure}[!p]
\centering
\subfigure[]
{\includegraphics[scale=0.2]{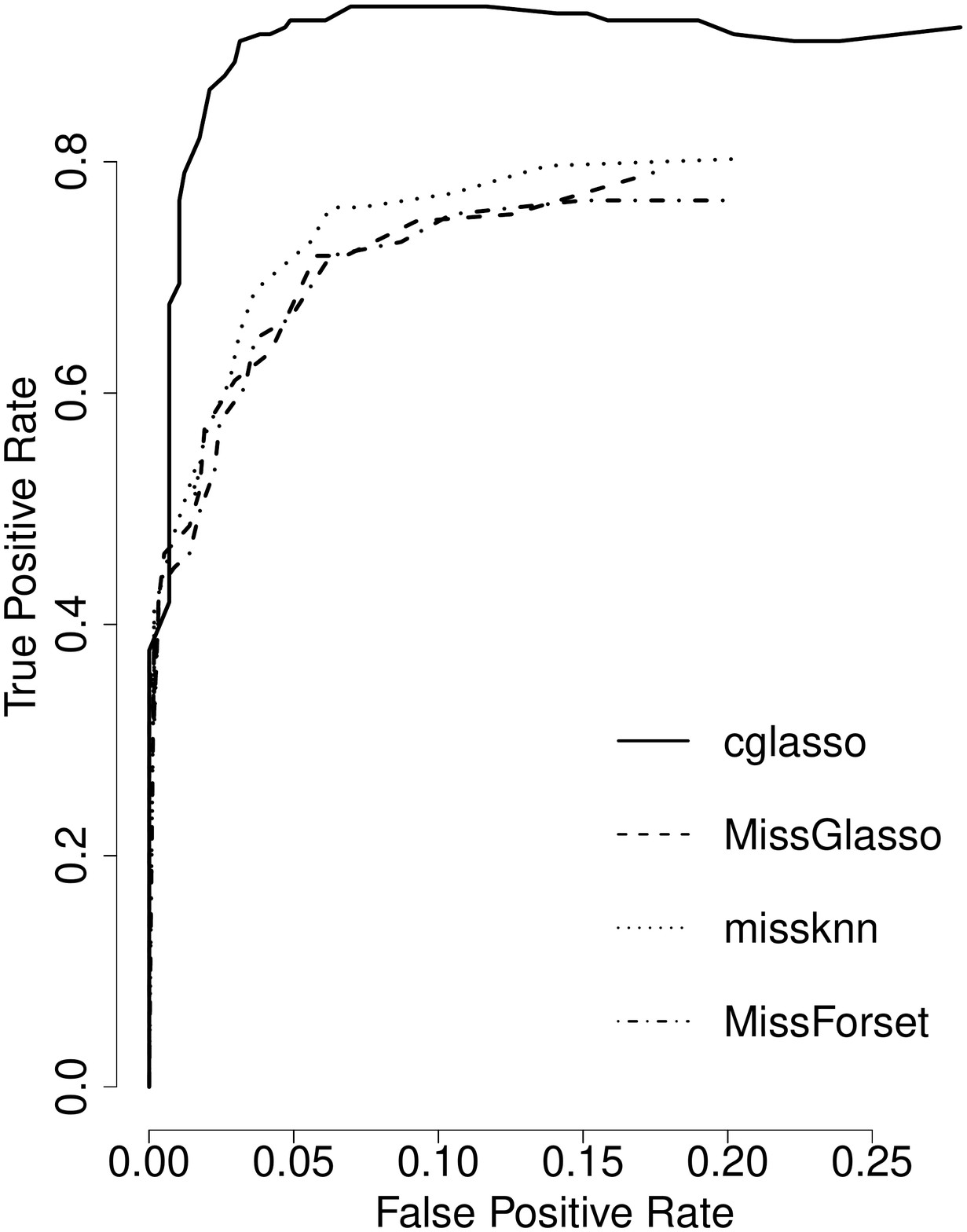}}
\subfigure[]
{\includegraphics[scale=0.2]{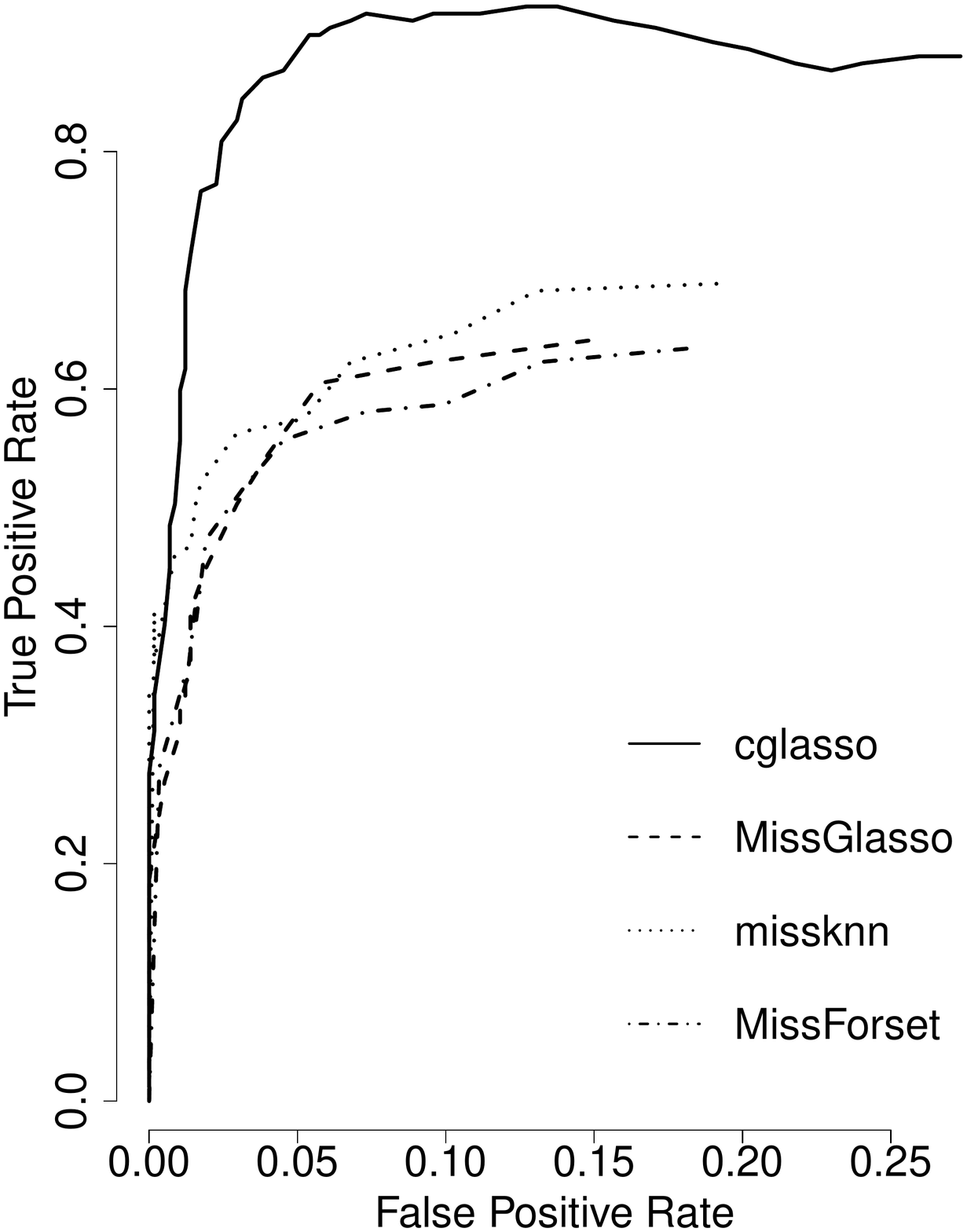}}
\subfigure[]
{\includegraphics[scale=0.2]{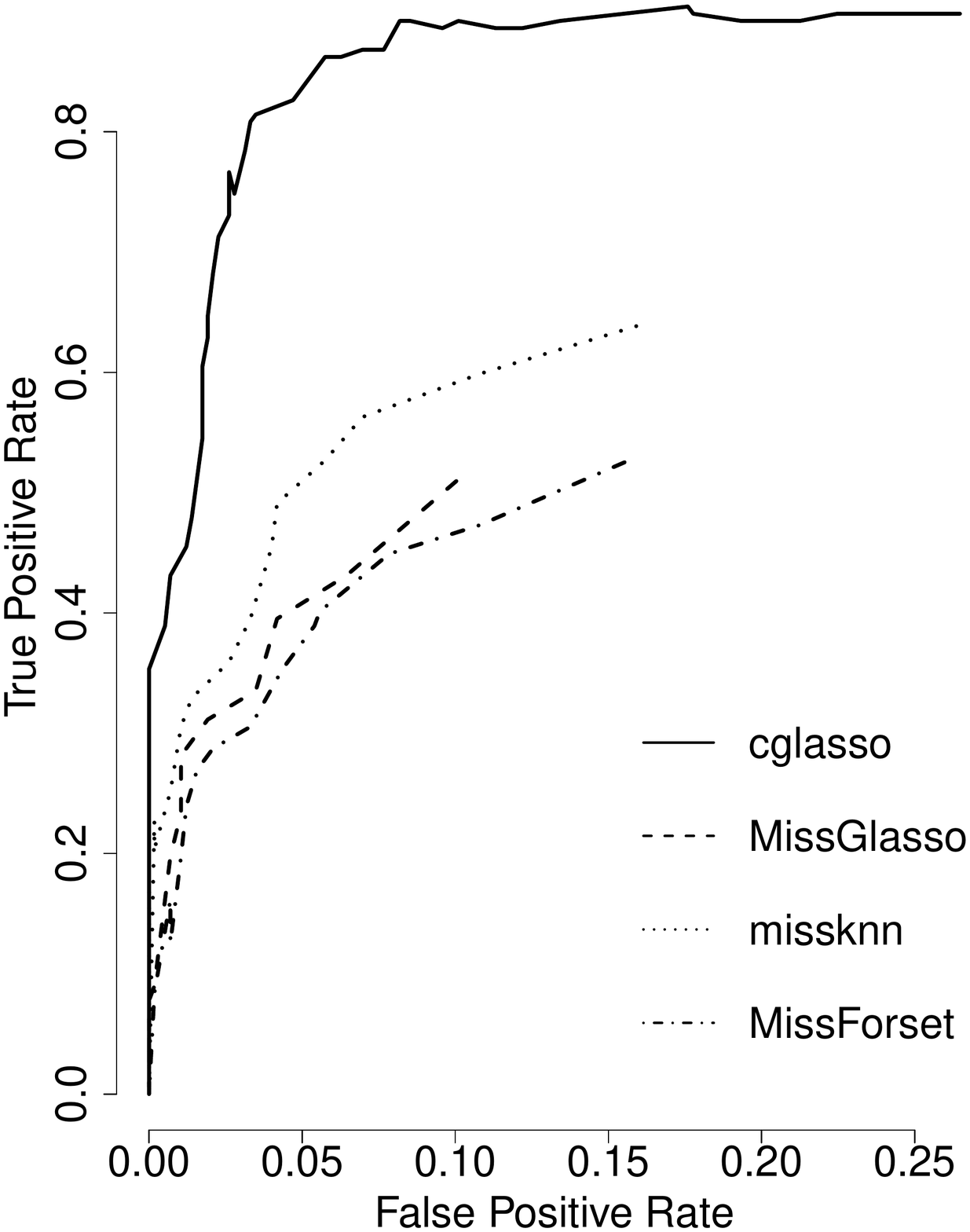}}
\caption{Comparison of methods on the Arabidopsis thaliana expression data: ROC curves comparing the network recovered by each method with that obtained using glasso on the fully observed data. Panel (a) reports the case of 10\% censoring, Panel (b) 20\%  and Panel (c) 30\%.}
\label{fig:censor}
\end{figure}

Table~\ref{tbl:ResCens} shows the Eucledian norm between the true observed data and the imputed one for the four methods considered and across the three different levels of censoring. The results show how the performance of all methods decreases the higher the level of censoring and how {\tt cglasso} is overall superior to the other methods across all comparisons. Figure~\ref{fig:censor} shows the ROC curves of the networks inferred by the four methods compared with the graphical lasso network from the fully observed data across the full path of penalisations and for different censoring levels. The proposed {\tt cglasso} method leads to a significant gain in network recovery over the existing methods: even with 30\% of censored data, the cglasso detects a similar network to that detected by glasso on the complete dataset.

\section{Application to single cell-data: blood development\label{sec:realdata}}

The increasing availability of single-cell expression data has led to the development of several statistical approaches aimed at gaining new insight into cell fate decisions and transitions between cell states. However, network inference from these data remains relatively unexplored aside from few recent examples (e.g., \citealp{ChanEtAl_CellSyst_17}; \citealp{OconeEtAl_BioInfo_15}). Recently, \citet{MoignardEtAl_NatBio_15} have conducted an in-vivo gene-expression analysis of early blood development at the single-cell level in order to reconstruct a transcriptional regulatory network model that can describe the molecular mechanisms of blood development. The authors focus on 33 transcription factors, involved in endothelial and haematopoietic development, and on 9 marker genes.  Using microfluidic RT-qPCR technology, the expression of these genes is measured on mouse embryonic cells across 1.25 days of post-implantation mouse development. For our analysis, we consider closely the last time point, where specific data are available on 770 endothelial cells which have a key role in blood development (\citealp{MoignardEtAl_NatBio_15}).

\begin{figure}[!p]
\centering
\subfigure[]
{\includegraphics[scale = 0.4]{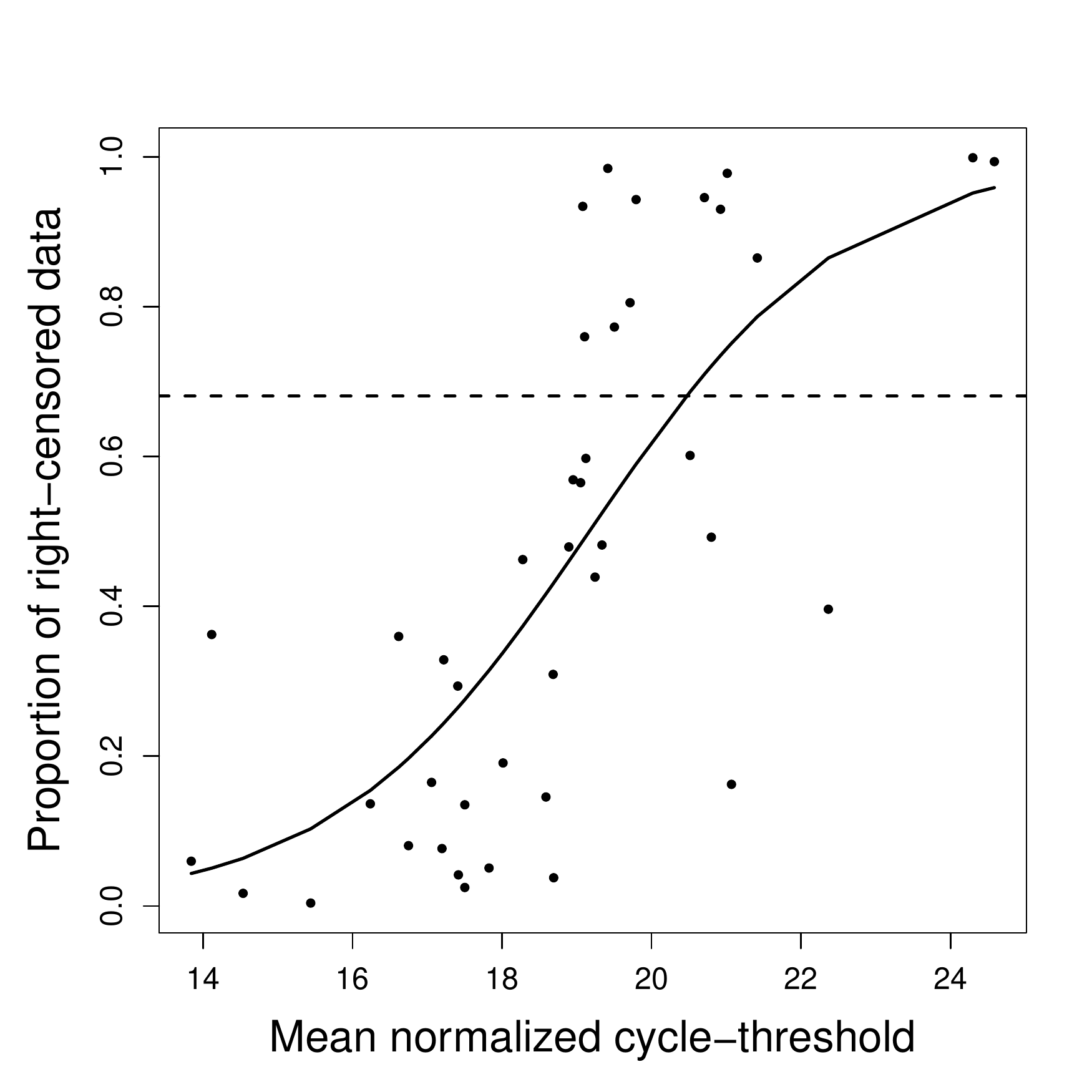}}
\subfigure[]
{\includegraphics[scale = 0.4]{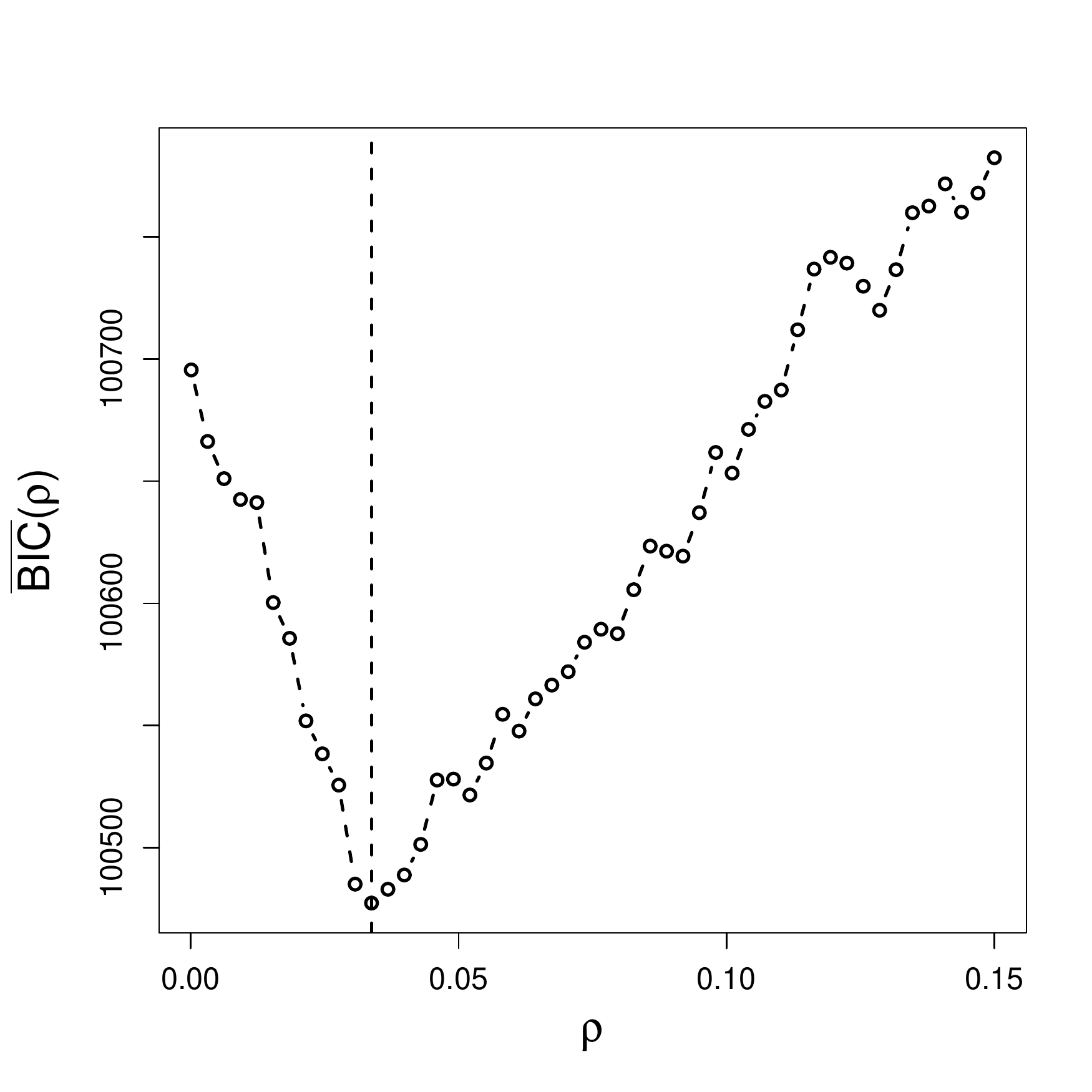}}\\
\subfigure[]
{\includegraphics[scale = 0.4]{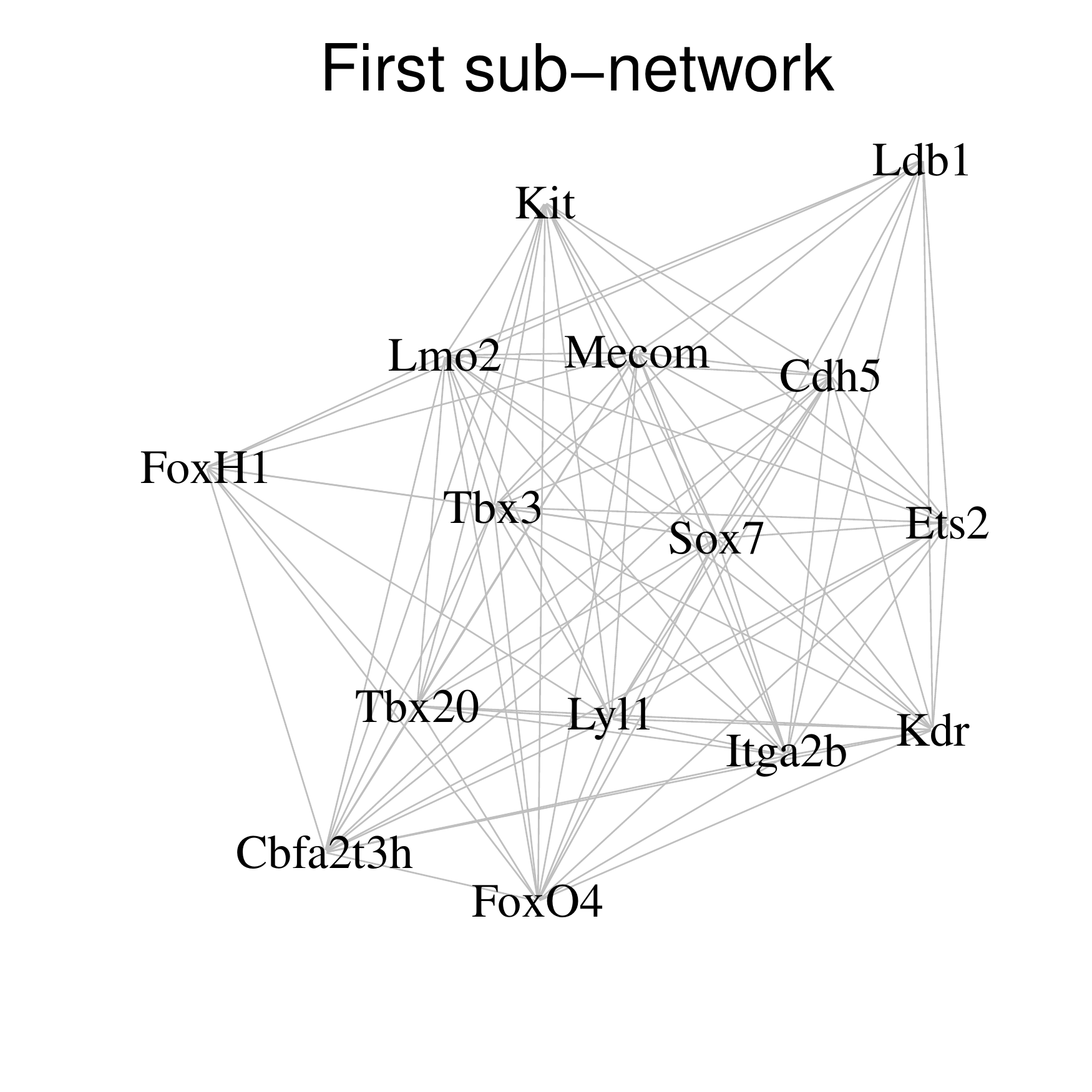}}
\subfigure[]
{\includegraphics[scale = 0.4]{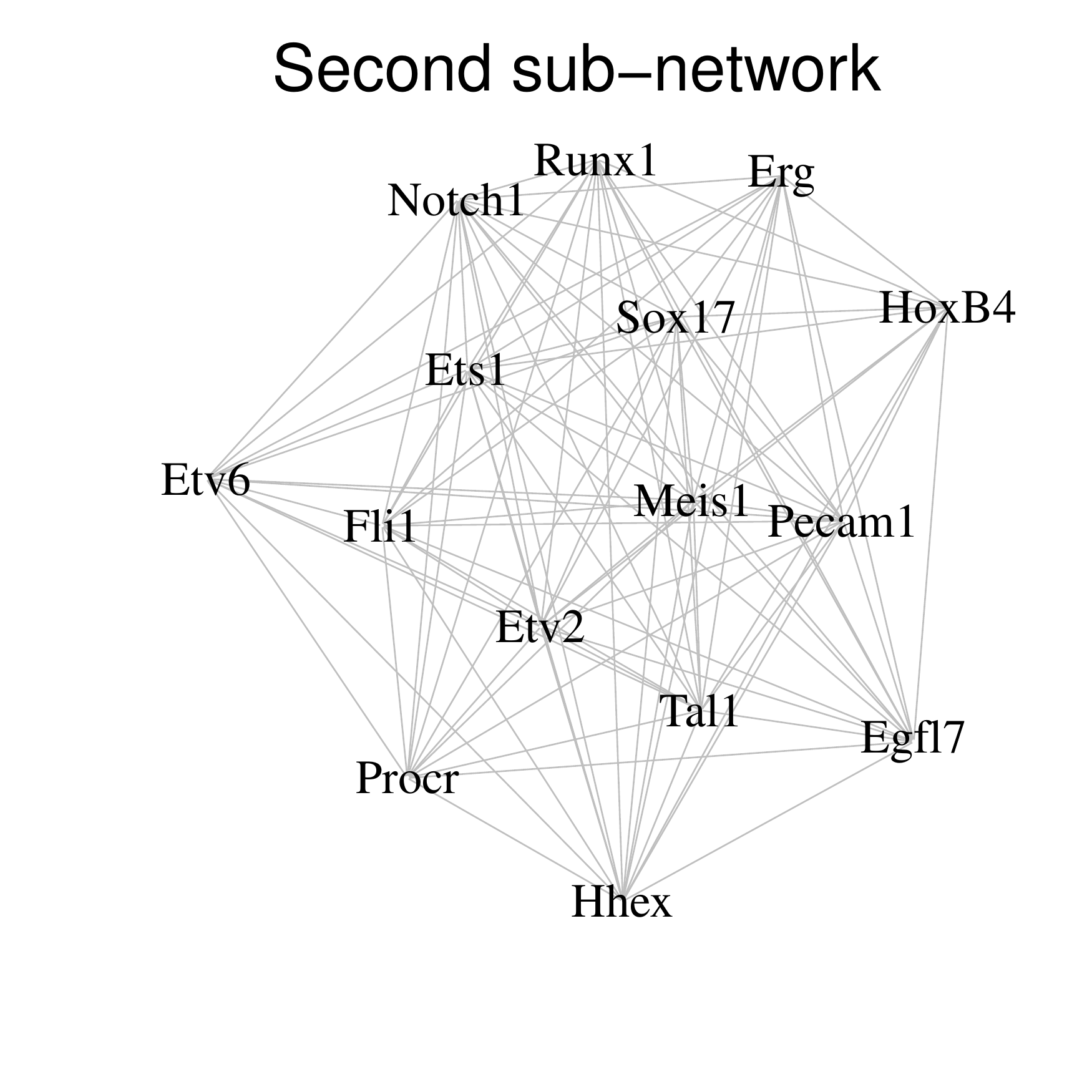}}
\caption{Panel (a): the proportion of right-censored data versus the mean of the normalized cycle-threshold; black line is obtained by fitting a logistic regression model while the dashed line identifies the threshold used to filter out the genes form the study. Panel (b): path of the $\overline{\hbox{BIC}}(\rho)$ measure; vertical dashed line identifies the optimal value of the tuning parameter. Panel (c): first sub-network. Panel (d): second sub-network.}
\label{fig:analysis_dataset}
\end{figure}

As discussed in the Introduction, RT-qPCR data are typically right-censored. In this particular study, the limit of detection is quantified using the method proposed in \citet{StahlbergEtAl_NatAcidRes_11} and manufacturer's instructions and the resulting right-censoring value is fixed to 25. Raw data are mean normalized applying the method proposed in \citet{PipelersEtAl_Plos1_17} and using the 4 housekeeping genes provided. Panel~(a) in  Figure~\ref{fig:analysis_dataset} shows the relationship between the proportion of right-censored data and the mean of the normalized cycle-threshold. Two main conclusions can be drawn from this figure. Firstly, the proportion of censored data is an increasing function of the mean of the normalized cycle-threshold: this is expected and means that the assumption of missing-at-random is not justified on this dataset. Secondly, there are some genes with a very high proportion of censoring (see the points above the dashed line): these are genes that are either not required at the considered time point of the blood development or whose transcription failed to amplify. Following this explorative analysis and considering also the possible computational problems caused by the inclusion of these genes, we filtered out the genes with a proportion of censoring above 70\% for subsequent analysis.

The normalized cyclic thresholds of the remaining  30 genes are used to fit a  right-censored GGM by using the proposed cglasso estimator. Panel~(b) in  Figure~\ref{fig:analysis_dataset} shows the path of the $\overline{\hbox{BIC}}(\rho)$ measure and the vertical dashed line is traced in correspondence of the optimal $\rho$-value. The resulting estimated concentration graph contains about $80\%$ of all possible edges. A community detection method based on modularity (\citealp{Newman_PHE_06}) identifies two clear sub-networks. These are depicted in Panel~(c) and (d), respectively. Among the most connected genes, we find endothelial genes that are well known in the literature, such as Erg, Sox7, Sox17, Hoxb4 and Cdh5. The last one is a known endothelial marker (\citealp{MoignardEtAl_NatBio_15}) while Erg is an essential transcription factor for definitive haematopoiesis and adult hematopoietic stem cell function. \citet{MoignardEtAl_NatBio_15} show that Erg expression is  activated by Sox17, while previous studies have shown that Erg is controlled by Ets factors and is active during haematopoietic development (\citealp{WilsonEtAl_Blood_09}) and in hematopoietic stem cells (\citealp{ThomsEtAl_Blood_11}). Finally, Sox17 belongs to the SoxF family of transcription factors which have recently been shown to confer arterial identity in combination with RBPJ/Notch (\citealp{SacilottoEtAl_PNAS_13}). The second sub-network also identifies previously known interactions, including close connections between Etv2, Fli1 and Tal1, where the latter two are known to function downstream of Etv2 in the haemangioblast.

\section{Conclusion\label{sec:conclusions}}
In this paper, we have proposed a definition for censored Gaussian Graphical models. The definition includes the standard Gaussian graphical model and the right and left censored graphical models as special cases. Since classical Gaussian graphical models cannot be used in high-dimensional settings, we also introduced $\ell_1$-penalisation to produce sparsity (model selection) and parameter estimation simultaneously. The resulting estimator is called censored graphical lasso (cglasso). The computational problem of estimating the mean and conditional independence graph of a censored Gaussian graphical model is solved via an EM-algorithm. We also derived an approximated EM-algorithm which is computationally more efficient and can be used for high-dimensional data analyses. An extensive simulation study showed that the proposed estimator overcomes the existing estimators both in terms of parameter estimation and of network recovery. The analysis of a real RT-qPCR dataset showed how the method is able to infer the regulatory network underlying blood development under high levels of censoring.

%\section*{Supplementary Materials}

%Supplementary material is available at \url{http://biostatistics.oxfordjournals.org}.

\section*{Funding}

The project was partially supported by the \emph{``European Cooperation in Science \& Technology''} (COST) funding: action number CA15109.

\bibliographystyle{chicago}
\bibliography{cglassobib}

\begin{table}[!p]
\caption{Pseudo-code of the proposed EM algorithm~\label{algo:cglasso1}}
%{\tabcolsep=4.25pt
\begin{tabular}{cl}
  \hline\\%
Step & Description \\ \hline
1. & Let $\{\bm{\hat\mu}^\rho_{ini},\widehat\Theta^\rho_{ini}\}$ be an initial estimate and $N$ the maximum number of steps\\
2. & for $n = 1$ to $N$ do\\
    & \multicolumn{1}{c}{\textbf{E-Step}}\\
3. & \qquad compute $\bm{\bar x}(\bm{\hat\mu}^\rho_{ini},\widehat\Theta^\rho_{ini})$ and $S(\bm{\hat\mu}^\rho_{ini},\widehat\Theta^\rho_{ini})$\\
    & \multicolumn{1}{c}{\textbf{M-Step}}\\
4. & \qquad $\bm{\hat\mu}^\rho \gets \bm{\bar x}(\bm{\hat\mu}^\rho_{ini},\widehat\Theta^\rho_{ini})$\\
5. & \qquad solve a standard graphical lasso problem using $S(\bm{\hat\mu}^\rho_{ini},\widehat\Theta^\rho_{ini})$:\\
    & \qquad $\widehat\Theta^\rho \gets \arg\max_{\Theta\succ0}\log\det\Theta - \mathrm{tr}\{\Theta S(\bm{\hat\mu}^\rho_{ini},\widehat\Theta^\rho_{ini})\} - \rho\sum_{h,k}|\theta_{hk}|$\\
6. & \quad if a a convergence criterion is met then\\
7. & \qquad return  $\{\bm{\hat\mu}^\rho,\widehat\Theta^\rho\}$\\
8. & \quad else\\
9. & \qquad $\bm{\hat\mu}^\rho_{ini}\gets\bm{\hat\mu}^\rho$ and $\widehat\Theta^\rho_{ini}\gets\widehat\Theta^\rho$\\
10. & end for 
\end{tabular}%}
\end{table}

\begin{table}[!p]
\caption{Pseudo-code of the proposed algorithm to compute the paths of the estimated parameters~\label{algo:cglasso2}}
%{\tabcolsep=4.25pt
\begin{tabular}{cl}
  \hline\\%
Step & Description\\ \hline
1. & Let $\rho_{\min}$ the smallest value of the tuning parameter\\
2. & Compute $\rho_{\max}$ as specified in Theorem~\ref{thm:maxrho}\\
3. & Compute a decreasing sequence of $K$ distinct $\rho$-values starting from $\rho_{\max}$ to $\rho_{\min}$, i.e.\\
    & \multicolumn{1}{c}{$\rho_{\max} = \rho_{(1)} > \cdots > \rho_{(K)} = \rho_{\min}$}\\
4. & for $k = 2$ to $K$ do\\
5. & \quad $\bm{\hat\mu}_{ini}^{\rho_{(k)}}\gets\bm{\hat\mu}^{\rho_{(k-1)}}$ and $\widehat\Theta^{\rho_{(k)}}_{ini}\gets\widehat\Theta^{\rho_{(k-1)}}$\\
6. & \quad Compute $\bm{\hat\mu}^{\rho_{(k)}}$ and $\widehat\Theta^{\rho_{(k)}}$ using Algorithm~\ref{algo:cglasso1} with $\{\bm{\hat\mu}_{ini}^{\rho_{(k)}}, \widehat\Theta^{\rho_{(k)}}_{ini}\}$ as initial estimates\\
7. & end for
\end{tabular}%}
\end{table}

\begin{table}[!p]
\caption{Results of the simulation study~\label{tbl:CompEff}}
%{\tabcolsep=4.25pt
\begin{tabular}{rccccccc}
  \hline\\%
& \multicolumn{7}{c}{Number of censored variables}\\ \hline
& 2 & 3 & 4 & 5 & 6 & 7 & 8\\
$\max_{\rho}\|\Delta\bm{\hat\mu}^\rho\|^2$ 	& 2.9$\times10^{-6}$ & 8.2$\times10^{-6}$ & 1.7$\times10^{-5}$ & 2.2$\times10^{-5}$ & 7.8$\times10^{-5}$ & 1.1$\times10^{-4}$ & 2.1$\times10^{-4}$ \\
												& (8.6$\times10^{-6}$) & (2.1$\times10^{-5}$) & (3.7$\times10^{-5}$) & (3.1$\times10^{-5}$) & (1.0$\times10^{-4}$) & (1.2$\times10^{-4}$) & (2.3$\times10^{-4}$)\\[2pt]
$\max_{\rho}\|\Delta\hat\Theta^\rho\|^2_F$ 	& 3.0$\times10^{-5}$ & 8.3$\times10^{-5}$ & 2.0$\times10^{-4}$ & 2.6$\times10^{-4}$ & 2.3$\times10^{-3}$ & 2.6$\times10^{-3}$ & 6.6$\times10^{-3}$\\
												& (8.5$\times10^{-5}$) & (2.0$\times10^{-4}$) & (3.7$\times10^{-4}$) & (3.8$\times10^{-4}$) & (3.1$\times10^{-4}$) & (2.4$\times10^{-3}$) & (6.0$\times10^{-3}$)
\end{tabular}%}
\end{table}

\begin{table}[!p]
\caption{Results of the simulation study: for each measure used to evaluate the behaviour of the considered methods we report average values and standard deviation between parentheses \label{tbl:ResSimul}}
%{\tabcolsep=4.25pt
\begin{tabular}{ccc|ccccc|ccc}
  \hline\\%
\multicolumn{3}{c}{Model} & \multicolumn{2}{c}{$\min_\rho\mbox{MSE}(\bm{\hat\mu}^\rho)$} & \multicolumn{3}{c}{$\min_\rho\mbox{MSE}(\widehat\Theta^\rho)$} & \multicolumn{3}{c}{AUC}\\ \hline
 $p$ & $H/p$ & $k/p$  & cglasso & MissGlasso & cglasso & glasso & MissGlasso& cglasso & glasso & MissGlasso \\
50 & 0.5 & 0.06	& 0.47	& 14.50	& 8.76	& 103.35	& 96.75	& 0.60	& 0.46 	& 0.37\\
	&		& 		& (0.11)	& (0.69)	& (0.64)	& (14.43)	& (16.01)	& (0.02)	& (0.02)	& (0.02) \\ \hline
50 & 0.7 & 0.06 	& 0.48	& 21.00	& 10.11	& 139.76	& 131.99	& 0.58	& 0.39	& 0.25 \\
	&		&		& (0.10)	& (0.76)	& (0.84)	& (15.94)	& (18.81)	& (0.02)	& (0.02)	& (0.02)\\ \hline
50 & 0.6 & 0.02 	& 0.47	& 18.31	& 6.92	& 128.60	& 119.65	& 0.66	& 0.53	& 0.39 \\
	&		&		& (0.10)	& (0.81)	& (0.82)	& (17.75)	& (17.38)	& (0.02)	& (0.02)	& (0.03)\\ \hline
50 & 0.6 & 0.10 	& 0.46	& 17.39	& 12.02	& 113.84	& 105.70	& 0.53	& 0.37	& 0.29 \\
	&		&		& (0.10)	& (0.92)	& (0.85)	& (14.79)	& (15.71)	& (0.02)	& (0.02)	& (0.02) \\ \hline
200 & 0.5 & 0.015 & 1.92	&  63.34	& 41.57	& 398.02	& 373.86	& 0.34	& 0.21	& 0.07 \\
	&		&		& (0.19)	& (1.54)	& (1.52)	& (29.78)	& (32.46)	& (0.01)	& (0.01)	& (0.01)
\end{tabular}%}
\end{table}

\begin{table}[!p]
\caption{Comparison of methods on the Arabidopsis thaliana expression data: Euclidian norm between the observed and imputed data across the four methods and the three levels of censoring.\label{tbl:ResCens}}
\begin{tabular}{ccccc}
  \hline\\%
 censoring & cglasso & missknn & MissForest & MissGlasso \\
  \hline
10\% & 9.80 & 31.37 & 33.36 & 26.86 \\
  20\% & 14.85 & 40.16 & 46.26 & 36.45 \\
  30\% & 19.38 & 50.08 & 57.38 & 47.16
\end{tabular}
\end{table}

\end{document}